\documentclass[twocolumn]{aastex701}
\usepackage{graphicx,amsmath}
\usepackage{mathtools}

\newcommand{\Teff}{\ensuremath{T_{\rm eff}}}

\newcommand{\vrot}{\ensuremath{v_{\rm rot}}}

\newcommand{\etar}{\ensuremath{\eta_{\rm R}}}

\newcommand{\fc}{\ensuremath{f_{\rm c}}}
\newcommand{\fmu}{\ensuremath{f_{\mu}}}

\newcommand{\set}[1]{\texttt{#1}}
\newcommand{\mesa}{\texttt{MESA}}
\newcommand{\MESA}{\mesa}
\newcommand{\rsp}{\texttt{RSP}}
\newcommand{\RSP}{\rsp}

\newcommand{\mdc}{\texttt{mesh\_delta\_coeff}}
\newcommand{\tdc}{\texttt{time\_delta\_coeff}}
\newcommand{\dhr}{\texttt{delta\_HR\_limit}}
\newcommand{\dhrt}{\texttt{delta\_lgTeff\_limit}}
\newcommand{\dhrl}{\texttt{delta\_lgL\_limit}}
\newcommand{\smp}{\texttt{num\_cells\_for\_smooth\_gradL\_composition\_term}}

\newcommand{\PL}{\ifmmode{P-L}\else$P-L$\fi}
\newcommand{\PR}{\ifmmode{P-R}\else$P-R$\fi}
\newcommand{\PA}{\ifmmode{P-{\rm Age}}\else$P-{\rm Age}$\fi}
\newcommand{\ML}{\ifmmode{M-L}\else$M-L$\fi}

\newcommand{\MS}{\ifmmode{\,}\else\thinspace\fi{\rm M}\ifmmode_{\odot}\else$_{\odot}$\fi}
\newcommand{\LS}{\ifmmode{\,}\else\thinspace\fi{\rm L}\ifmmode_{\odot}\else$_{\odot}$\fi}
\newcommand{\RS}{\ifmmode{\,}\else\thinspace\fi{\rm R}\ifmmode_{\odot}\else$_{\odot}$\fi}
\newcommand{\feh}{\ifmmode{\rm [Fe/H]}\else[Fe/H]\fi}
\newcommand{\fcor}{\ifmmode{f_{\rm H}}\else$f_{\rm H}$\fi}
\newcommand{\fenv}{\ifmmode{f_{\rm env}}\else$f_{\rm env}$\fi}
\newcommand{\fhe}{\ifmmode{f_{\rm He}}\else$f_{\rm He}$\fi}

\newcommand{\npg}{\ifmmode{^{14}\rm N (\rm p,\gamma) ^{15} \rm O}\else{$^{14}\rm N (\rm p,\gamma)^{15}\rm O$}\fi}
\newcommand{\cag}{\ifmmode{^{12}\rm C (\alpha,\gamma) ^{16} \rm O}\else{$^{12}\rm C (\alpha,\gamma)^{16}\rm O$ }\fi}

\defcitealias{Ziolkowska-2024}{P1}
\defcitealias{Ziolkowska-2026}{P2}
\defcitealias{Smolec-2026}{P3}

\begin{document}

\title{Toward a Comprehensive Grid of Cepheid Models with MESA. IV. Modest Effects of Rotation on Blue Loops}

\correspondingauthor{Radoslaw Smolec}
\author[orcid=0000-0001-7217-4884,sname='Smolec']{Radoslaw Smolec}
\affiliation{Nicolaus Copernicus Astronomical Centre, Polish Academy of Sciences, Bartycka 18, 00-716 Warszawa, Poland}
\email[show]{smolec@camk.edu.pl}

\author[orcid=0000-0002-7448-4285,sname='Rathour']{Rajeev Singh Rathour}
\email{rajeevsr@camk.edu.pl}
\affiliation{Université C\^{o}te d'Azur, Observatoire de la C\^{o}te d'Azur, CNRS, Laboratoire Lagrange, France}

\author[orcid=0000-0002-3643-0366,sname='Hocd\'{e}']{Vincent Hocd\'{e}}
\email{vhocde@camk.edu.pl}
\affiliation{Nicolaus Copernicus Astronomical Centre, Polish Academy of Sciences, Bartycka 18, 00-716 Warszawa, Poland}

\begin{abstract}
Evolutionary tracks for 2--8\MS{} stars, with metallicities of $Z=0.014$, $0.006$, and $0.002$, including rotation, are computed with Modules for Experiments in Stellar Astrophysics (\MESA). We study how rotation impacts the evolutionary properties of classical Cepheids. We examine whether rotation can offer a plausible explanation for the mass discrepancy problem when it is included in the evolutionary code using the fully diffusive approximation for rotationally induced mixing processes. We find that rotation barely influences the appearance and luminosity levels of the blue loops. While luminosity increases with increasing initial rotation rate, the increase does not exceed 0.04\,dex, a fraction of the increase resulting from including the main sequence (MS) core overshooting of $0.2H_p$. As a consequence, rotation alone cannot resolve the mass discrepancy problem without simultaneously requiring significant MS core overshooting. Similar to the mass–luminosity relation, the period–radius and period–luminosity relations are barely affected by rotation, while the period–age relation predicts Cepheid ages to be only a few per cent longer compared with models without rotation. The predicted surface rotational velocities are too large compared with observations. These results are in contrast with those obtained with the Geneva code, which implements rotational mixing using the advective–diffusive scheme. In that approach, the luminosity levels of the loops are significantly higher, their luminosity extent increases, and the predicted rotation velocities are lower, compared with \MESA{} models. The differences between the two approaches arise from significantly more efficient rotation-induced mixing during the MS evolution in models computed with the advective–diffusive scheme.
\end{abstract}



\section{Introduction}

Classical Cepheids are one of the most important classes of variable stars in astrophysics, forming an essential rung on the cosmic distance ladder \citep[e.g.,][]{Freedman-2001, Riess-2022}. This role stems from the period-luminosity relation (\PL), the Leavitt law \citep{Leavitt-1912}. However, their role is not limited to standard candles. Thanks to the period-age relation (\PA), they are used in galactic archaeology \citep[e.g.,][]{Jacyszyn-2016, Skowron-2019, DeSomma-2025}. As evolutionarily advanced stars -- most of them burn helium in their cores -- they allow us to test both evolutionary and pulsation theories. This is achieved, among other things, by the period-radius (\PR) and mass-luminosity (\ML) relations. All these relations are a consequence of Cepheids occupying a well-defined place on the Hertzsprung-Russell diagram (HRD), within the classical instability strip (IS), where Cepheids exhibit high-amplitude radial pulsations. Three IS crossings are possible: the first (1st) during the relatively rapid evolution after the main sequence (MS), when the star moves toward the red giant branch (RGB), and the second (2nd) and third (3rd) during core helium burning, when the star traces an almost horizontal blue loop on the HRD.

The study of the \ML{} relation reveals one of the unsolved puzzles of Cepheids -- the mass discrepancy problem. Their masses, determined on the basis of evolutionary calculations, are too large compared with the masses determined from pulsation calculations \citep{Stobie-1969, Cox-1980}. After revising the opacity tables \citep[see][]{IglesiasRogers-1991, Moskalik-1992}, the discrepancy remains at about 20\% \citep{Keller-2008}. Direct measurements of Cepheid masses in eclipsing binary systems show that the pulsation theory correctly predicts the Cepheid masses, and the solution should be sought in the theory of evolution \citep[see][]{Pietrzynski-2010, Pilecki-2018}.

Possible solutions to the problem include incorporating processes into the models that increase the luminosity for a given Cepheid mass. These are the overshooting from the MS convective core \citep[e.g.,][]{CassisiSalaris-2011, PradaMoroni-2012} and rotation \citep[][]{Anderson-2014, Anderson-2016, Miller-2020}. Another process under consideration is the pulsation-induced mass loss, which reduces the Cepheid mass on the blue loop \citep[e.g.,][]{Neilson-2011, Neilson-2012}. For a review of the possibilities, see \cite{Bono-2006}.

In this paper, we investigate the role of rotation on the \ML{}, \PL{}, \PR{}, and \PA{} relations using evolutionary tracks computed with the Modules for Experiments in Stellar Astrophysics, \MESA{} \citep{Paxton-2011, Paxton-2013, Paxton-2015, Paxton-2018, Paxton-2019, Jermyn-2023}.

Classical Cepheids are considered slow rotators. Unfortunately, it is very difficult to measure their rotation rates, as it is not straightforward to disentangle the various contributions to spectral line widths, which include turbulent motions in the atmosphere, large-amplitude radial oscillations, and rotation. \cite{BersierBurki-1996}, by an analysis of line broadening for 41 Cepheids, concludes that the equatorial rotational velocity is smaller than 10 km/s. Based on spectroscopic observations of nine Cepheids, \cite{Nardetto-2006} provides the following pulsation-period-dependent relation for the projected velocity, $V_{\rm rot}\sin i = (-11.5\pm0.9)\log P + (19.8\pm1.0)$ [km/s]. Based on this relation, the rotation velocities of classical Cepheids should typically fall in the $\sim10-25$\,km/s range. \cite{Anderson-PhD} gives an average equatorial rotational velocity for a sample of 97 MW Cepheids of 12.3\,km/s. Significantly larger velocities are excluded, as they would manifest in a significant broadening of spectral lines, which is not observed. While Cepheids rotate slowly, their progenitors on the MS, the B-type stars, are relatively fast rotators, with projected rotational velocities of the order of 100--150\,km/s \citep{Huang-2010}. Their early evolution leaves a significant imprint on all the subsequent phases, including the core helium burning.

\cite{Anderson-2014, Anderson-2016} analyzed the Geneva evolutionary tracks including rotation, computed by \cite{Georgy-2013}, in the context of the classical Cepheids and their mass discrepancy problem. They considered three metallicities, $Z=0.014$, $0.006$, and $0.002$, typical for the Milky Way (MW) and the Large and Small Magellanic Clouds (LMC and SMC), respectively. They found that rotation significantly impacts the \ML{} relation. The luminosity level of the blue loop not only increases, but also the luminosity separation between the 2nd and 3rd crossings ({\it the thickness} of the loop) increases considerably. They argue that, by taking into account rotation and the crossing number, the Cepheid mass discrepancy vanishes, without the need to introduce the high MS core overshooting (they adopt $0.1H_p$), or the pulsation-driven mass loss. They also note a significant impact of rotation on the \PA{} relation, with the rotating models predicting significantly ($\sim$20\%) older Cepheids, compared with the relations based on the non-rotating models.

The just described effects of rotation have their origin in the rotation-induced transport of angular momentum and in the rotation-induced mixing of chemical elements -- for reviews, see e.g., \cite{MaederMeynet-2000}, \cite{Maeder-2009book}, and \cite{Aerts-2019-AMRev}. There is no unique way to include rotation and the associated mixing processes in the stellar evolution codes. Two main approaches are used in the literature: the advective–diffusive scheme, in which the transport of angular momentum due to meridional circulation is treated advectively while the remaining mixing processes are described in the diffusion approximation, and the fully diffusive scheme. A more detailed discussion will be presented in Sect.~\ref{ssec:methods_rotation}. The advective-diffusive scheme is implemented in the Geneva code, while the diffusive scheme is implemented in \MESA. It is now well established that these schemes may give discrepant results \citep[e.g.,][]{Choi-2016, Martinelli-2025}. The discrepancies for the classical Cepheid models were discussed by \cite{EspinozaArancibia-2022}, who used \MESA{} and models including rotation to study the pulsation period change rates\footnote{We note that their comparison with the Geneva tracks is flawed, as they did not notice the different definitions of the critical angular velocity in the \MESA{} and Geneva codes, see Sect.~\ref{ssec:oocrit}.}. The major goal of this paper is to systematically study the effects of rotation, as implemented in \MESA, on the evolutionary tracks of Cepheids and on their properties, including the \ML, \PL, \PR, and \PA{} relations, and to conduct a comparison with the results obtained with the advective-diffusive scheme, as implemented in the Geneva code.

This paper is the fourth in a series devoted to the investigation of the evolutionary and pulsation properties of the classical Cepheid models computed with \MESA. In \citet{Ziolkowska-2024} (hereafter \citetalias{Ziolkowska-2024}), we established the numerical and microphysical setup for the reference models and quantified the uncertainties of the evolutionary tracks arising from different choices of physical ingredients (e.g., Mixing Length Theory (MLT), nuclear reaction rates, the reference solar composition). In \citet{Ziolkowska-2026} (hereafter \citetalias{Ziolkowska-2026}), we investigated the corresponding uncertainties in the predicted surface abundances. In \citet{Smolec-2026} (hereafter \citetalias{Smolec-2026}), we computed an extensive grid of the evolutionary models covering the $2-8\MS$ mass range and metallicities from $\feh=-1.0$ to $+0.2$, exploring the effects of the core and envelope overshooting, mass loss, nuclear reaction rates, and the adopted solar mixture on the morphology of the blue loops and on the Cepheid pulsation properties. Based on the evolutionary and pulsation calculations, we derived the $\PL$, $\ML$, $\PR$, and $\PA$ relations for the fundamental-mode (F mode) Cepheids and examined their dependence on metallicity. In the present paper, we study the impact of rotation on the evolutionary properties of the Cepheids.

The paper is organized as follows. In Sect.~\ref{sec:methods}, we present the methods behind the evolutionary and pulsation calculations, with a focus on the rotational setup of the models. In Sect.~\ref{sec:results}, we present our results, in particular discuss the impact of rotation on the appearance of the blue loops and on the evolutionary and pulsation relations. In Sect.~\ref{sec:comparison}, we conduct a comparison with other tracks incorporating rotation, most notably with the Geneva tracks. We conclude the paper with a summary in Sect.~\ref{sec:summary}.

\section{Methods\label{sec:methods}}

\subsection{Evolutionary calculations}\label{ssec:methods_evol}

In all our evolutionary calculations, we use \MESA, version r-21.12.1. In \citetalias{Smolec-2026}, we have demonstrated that the differences when computing the non-rotating models with a more recent version of \mesa{}, 24.05.01, are negligible, except for the solar metallicity models without overshooting, the minor differences recorded there being likely due to slightly different sets of nuclear reactions in the two \MESA{} versions. In Sect.~\ref{secapp:oldvsnew} in the Appendix, we repeat a similar exercise with \MESA{} version 25.12.1\footnote{This was the most recent release at the time of writing. While the paper was in review, version 26.4.1 was released}, and the models including rotation, again finding no significant differences. We use r-21.12.1 for consistency with our earlier work, \citetalias{Ziolkowska-2024}--\citetalias{Smolec-2026}.

The microphysical setup of our models is exactly the same as in the reference model discussed in \citetalias{Ziolkowska-2024} and in the \citetalias{Smolec-2026} models. Below we summarize the most important information.

All our models adopt the OPAL opacity tables \citep{Iglesias-1993, Iglesias-1996} at lower temperatures supplemented with the \cite{Ferguson-2005} opacity data. The \cite{Asplund-2009} (A09) reference solar composition is adopted. For the helium enrichment, we assume $\Delta Y/\Delta Z=1.5$. With the primordial helium abundance of $Y_{\rm p}=0.2485$ \citep{Komatsu-2011}, the three metallicities we consider in this study, $Z=0.014$, $Z=0.006$, and $Z=0.002$, correspond to the helium mass fractions of $Y=0.2695$, $Y=0.2575$, and $Y=0.2515$, respectively, and the metallicities of $\feh{}=+0.03$, $-0.35$, and $-0.83$, respectively, see also tab.~1 in \citetalias{Smolec-2026}.

The standard \mesa{} equation of state is included, which is a blend of the OPAL \citep{Rogers2002}, SCVH \citep{Saumon1995}, FreeEOS \citep{Irwin2004}, HELM \citep{Timmes2000}, PC \citep{Potekhin2010}, and Skye \citep{Jermyn2021} EOSes.

The \mesa{} built-in \texttt{pp\_and\_cno\_extras.net} nuclear reaction network, which follows the evolution of 25 isotopes, is used in our calculations. The nuclear reaction rates come from NACRE \citep{nacre}. For the \cag{} reaction, the \cite{Kunz-2002} reaction rate is used, and for the slowest reaction in the CNO cycle, \npg{}, the JINA REACLIB rate is adopted \citep{jina}.

At the outer boundary, we use the \MESA's built-in PHOENIX atmosphere tables \citep{Hauschildt-1999a, Hauschildt-1999b} and the \cite{Castelli-2003} models, see \cite{Paxton-2011}.

The Mixing Length Theory (MLT) follows the formulation by \cite{Henyey1965} with $\alpha_{\rm MLT}=1.77$, resulting from the solar calibration performed in \citetalias{Ziolkowska-2024}.

Convective boundaries are set with the Schwarzschild criterion and the predictive mixing scheme \citep{Paxton-2018} at the core boundary. Overshooting follows the exponential prescription \citep{Herwig-2000}, with the efficiency characterized with the \fcor{} and \fenv{} parameters at the MS core and at the lower boundary of the convective envelope on the RGB, respectively. The overshooting may also be included with the step formalism, in which the extent of the overshooting region is characterized with a fraction of the local pressure scale height, $\beta H_p$. In \citetalias{Smolec-2026} (Sect.~A), we have derived the relation between the $\beta$ and $\fcor$ parameters, which depends little on the metallicity, but is somewhat sensitive to the mass. For rough comparisons, the $\beta=11.5\fcor$ relation may be used.

Following our earlier work, \citetalias{Ziolkowska-2024}–\citetalias{Smolec-2026}, overshooting from the helium-burning core is not included. Its inclusion may lead to numerical instabilities, as investigated by \cite{Ostrowski-2021}. While defining our reference model in \citetalias{Ziolkowska-2024}, we also checked that the inclusion of core-helium overshooting has a negligible effect on the properties of the blue loops. In Appendix~\ref{secapp:heov}, we also confirm that this is the case for models including rotation. The only significant effect of including core-helium overshooting is the appearance of secondary loops, while the overall properties of the blue loops, their extent, and luminosity level remain the same as in models without overshooting. Also, the surface rotational velocities are not affected.

The models are computed from ZAMS until the end of the core helium burning. The initial AGB ascent is also followed; this phase is included only for reference and better visualization of the evolutionary tracks, and is not modeled in detail.

Our models do not include the atomic diffusion.
The rotational setup and the treatment of mass loss are described in the following section.

\subsection{Rotation setup}\label{ssec:methods_rotation}

The details of the implementation of rotation in \MESA{} are described in \cite{Paxton-2013} and \cite{Paxton-2019}. \MESA{} adopts the shellular approximation, which is appropriate in the regime of the strong anisotropic turbulence induced by the differential rotation, which efficiently suppresses the compositional and velocity gradients along the isobars \citep{Zahn-1992, MeynetMeader-1997}. In this approximation, the structure equations are formulated on the isobars, with the centrifugal acceleration terms included through the correction factors, $f_p$ and $f_T$, which account for the distortion of the Roche potential from the spherical symmetry \citep{KippemhahnThomas-1970, EndalSofia-1976}. \cite{Paxton-2019} improved the treatment, particularly for the close-to-critical rotation, by replacing the numerical evaluations of these geometric correction factors with the analytic fits to the properties of the Roche potential of a point mass.

The transport of chemical elements and angular momentum by rotationally induced instabilities is treated within the diffusive approximation. In this respect, \MESA{} follows evolutionary codes like Kepler \citep{Heger-2005} or STERN \citep{Heger-2000, Brott-2011}, from which \MESA{}'s implementation of rotation was derived. The rotationally induced mixing processes that are included in the diffusion coefficient are Eddington–Sweet (meridional, ES) circulation, dynamical shear instability (DSI), Solberg–Høiland (SH) instability, secular shear instability (SSI), and the Goldreich–Schubert–Fricke (GSF) instability; see \cite{Heger-2000} for a detailed description of these instabilities and the computation of the corresponding diffusion coefficients. In \MESA, each of these diffusion coefficients is multiplied by a factor of order unity, which we denote $f_{\rm ES}$ for the meridional circulation, $f_{\rm DSI}$ for the dynamical shear instability, and so on. For each of these instabilities, different factors may be used in the angular momentum transport and chemical element mixing equations. In this study, we always use the same values in both equations (which is also the \mesa{} default). In the diffusion equation for chemical element mixing, the total diffusion coefficient arising from the rotational instabilities is multiplied by the $\fc$ factor, which scales the efficiency of chemical element mixing relative to that of angular momentum transport. Following \cite{ChaboyerZahn-1992} and \cite{Heger-2000}, we set $\fc=1/30$, but in Sect.~\ref{ssec:freep} we also explore other values. Another free parameter in the model is $\fmu$, which multiplies $\nabla_\mu$, i.e., it describes the sensitivity of the rotationally induced mixing to composition gradients. Following \cite{Heger-2000}, we adopt $\fmu=0.05$, but in Sect.~\ref{ssec:freep} we also explore the effects of other choices.

While \MESA{} includes magnetic angular momentum transport via the Spruit–Tayler dynamo \citep[ST;][]{Tayler-1973, Spruit-2002}, by default, we neglect magnetic fields and the associated instabilities in this work. The efficiency and robustness of this mechanism in stellar radiative zones remain debated \citep[e.g.,][]{Zahn-2007, Aerts-2019-AMRev, Fuller-2019}. In Sect.~\ref{ssec:freep}, we show its effects for a single set of models. When the contribution from the ST is included, the corresponding diffusion coefficient is computed following \cite{Heger-2005}, with scaling factors $f_{\rm ST}$ and $f_{\rm ST, AM}$ in the equations for chemical element and angular momentum mixing, respectively.

While the diffusive approximation is commonly adopted for the transport of chemical elements, for the angular momentum transport, the advective–diffusive scheme \citep{Zahn-1992, MeynetMeader-1997, MaederZahn-1998} is also used in stellar evolution calculations, for example in the Geneva code \citep{Ekstrom-2012}, FRANEC \citep[][which adopts both purely diffusive and advective–diffusive schemes]{ChieffiLimongi-2013}, or in STAREVOL \citep[see][]{Lagarde-2012, Amard-2019}. This formalism explicitly accounts for the large-scale meridional circulation as an advective process, while retaining diffusive terms to represent other instabilities. The advective–diffusive scheme provides a more physically consistent description of the angular momentum transport \citep{MaederMeynet-2000, Maeder-2009book}. Nevertheless, the diffusive approximation remains widely used due to its relative simplicity and numerical robustness. Different aspects of the two approaches were compared, e.g., in \cite{Choi-2016, MartinsPalacios-2013, EspinozaArancibia-2022, Martinelli-2025}. In Sect.~\ref{ssec:compgeneva}, we also compare our results (diffusive approach) with those obtained with the Geneva code.

For mass loss, we adopt similar options as \cite{Choi-2016}, allowing for mass loss already on the MS, where the `Dutch' wind scheme is adopted with the scaling parameter $\eta_{\rm Dutch}=1$. The Dutch wind is a mixture of prescriptions from \cite{NugisLamers-2000} and \cite{Vink-2001} -- see \cite{Glebbeek-2009}. For cool winds at later evolutionary stages, we use the \cite{Reimers-1975MSRSL} formula, with the scaling parameter $\eta_{\rm R}$ in the $0.4-0.8$ range, and later, on the AGB, the \cite{Bloecker-1995} prescription with $\eta_{\rm B}=0.2$. Mass loss is rotationally enhanced by a factor $(1-\Omega/\Omega_{\rm crit})^{-\zeta}$ \citep{FriendAbbott-1986, BjorkmanCassinelli-1993}, where we use $\zeta=0.43$ \citep{Langer-1998}. For the close-to-critical rotation, the implicit wind scheme is used, see \cite{Paxton-2013}.

For evolutionary calculations, in particular including diffusive processes, it is pertinent to conduct a convergence study to ensure that the results do not depend on the parameters controlling the spatial and temporal resolution of the models. For the non-rotating models, we have tested the convergence in \citetalias{Ziolkowska-2024}. With the same numerical resolution parameters, the rotating models are not converged; a higher spatial and temporal resolution is needed. The convergence study and the resulting parameters are presented in Sect.~\ref{secapp:convergence} in the Appendix.

\subsection{Critical angular velocity}\label{ssec:oocrit}

The models are initialized with a solid-body rotation just before reaching ZAMS, with a fraction of the critical angular velocity
\begin{equation}
W\coloneqq\Omega/\Omega_{\rm crit, M}. \label{eq:oocritmesa}
\end{equation}
In \MESA{}, the critical angular velocity is defined as \citep{Paxton-2013}:
\begin{equation}
\Omega_{\rm crit, M}^2=\left(1-\frac{L}{L_{\rm Edd}}\right)\frac{GM}{R^3}\,,
\end{equation}
where $L_{\rm Edd}$ is the Eddington luminosity and $L$, $M$, and $R$ are the luminosity, mass, and radius of the star, respectively. In intermediate-mass stars (with luminosities $\sim$2 orders of magnitude below the Eddington limit), the critical angular velocity may be written as
\begin{equation}
\Omega_{\rm crit, M}^2=\frac{GM}{R_{\rm e}^3}\,,\label{eq:omcritM}
\end{equation}
where $R_{\rm e}$ is the equatorial radius of the star. The above expression corresponds to the orbital velocity at the equator. Another definition of the critical velocity is adopted in the Geneva code, which corresponds to the critical breakup velocity at the equator \citep{Maeder-2009book, Ekstrom-2012}:
\begin{equation}
\Omega_{\rm crit, G}^2=\frac{GM}{R_{\rm e,crit}^3}\,,\label{eq:omcritG}
\end{equation}
where $R_{\rm e,crit}$ is the equatorial radius of the star rotating at the critical velocity.

Rotation rates, expressed with respect to these two critical rotation rates, $W$, eq.~\eqref{eq:oocritmesa}, and
\begin{equation}
 \omega\coloneqq\Omega/\Omega_{\rm crit, G}\,,\label{eq:oocritgeneva}
\end{equation}
are related by \citep{Rivinius-2013} (see also \cite{Martinelli-2025}):
\begin{equation}
 \omega=\frac{3\sqrt{3}W}{(2+W^2)^{3/2}}\,,\label{eq:omcritMG}
\end{equation}
and are the same only for non-rotating and critically rotating stars. Thus, for the same fraction of the critical velocity expressed through eqs.~\eqref{eq:omcritM} and \eqref{eq:omcritG}, we obtain significantly higher equatorial velocities with the definition adopted in \MESA. Since we will be comparing our results with those obtained with the Geneva code, we note that their $\omega_0=0.5$ and $\omega_0=0.9$ correspond to $W_0\approx 0.29$ and $W_0\approx 0.65$, respectively.

\subsection{Pulsation calculations}\label{ssec:puls}

Just as in \citetalias{Smolec-2026}, we use \mesa's Radial Stellar Pulsation module, \rsp{} \citep{Paxton-2019, sm08a}, to compute the linear pulsation properties of models along the stellar evolutionary tracks. \RSP{} constructs its own chemically homogeneous envelope model. The input parameters -- mass, effective temperature, absolute luminosity, and helium and metal mass fractions -- are taken directly from the evolutionary tracks. The microphysical ingredients adopted for the construction of the envelope and for the pulsation calculations (opacities and the equation of state) are exactly the same as those used in the evolutionary calculations. The numerical parameters defining the envelope model grid are also identical to those adopted in \citetalias{Smolec-2026}, to which we refer the reader for further details.

In this study, we focus exclusively on fundamental-mode properties. Accordingly, in the relations presented in the following sections, we use the linear fundamental-mode periods, $P_0$, computed with \RSP. Linear pulsation periods are sufficient for constructing the \PL{}, \PR{}, and \PA{} relations, as these depend primarily on the stellar structure (see Sect.~2.3 of \citetalias{Smolec-2026}).

\RSP{} does not include rotation, neither in the construction of the static model nor in the linear stability analysis. For radial modes, rotational effects enter only at the second order in the angular velocity and are therefore expected to be small for moderate rotation rates. The equilibrium models are constructed from the evolutionary tracks that include rotation, so the effects of rotation on global stellar parameters are incorporated indirectly. We note that this is the same approach as adopted in \cite{Anderson-2016}.

An advantage of \RSP{}, compared, for example, with \texttt{gyre} \citep{Townsend-gyre}, is that it includes time-dependent convection--pulsation coupling following the model of \cite{Kuhfuss-1986}. Consequently, linear analysis accounts for perturbations of the convective fluxes, allowing reliable growth rates -- and hence the location of the IS -- to be computed. In \citetalias{Smolec-2026}, we have studied in detail the dependence of the location of the IS on the physical parameters of the model, showing that the IS location weakly depends on the underlying $\ML$ relation of the models or on the crossing numbers. The strongest sensitivity was recorded for the parameters describing the convection–pulsation coupling, which resulted in the definition of the two fiducial ISs, hot and cool -- see Sect.~3.2 in \citetalias{Smolec-2026} for details. For consistency with our previous study, we adopt the hot IS as our default. The model properties will also be derived along the midline of the IS. Analytical expressions for the IS edges and the midline are given in Tab.~5 in \citetalias{Smolec-2026}.

\subsection{Grid of evolutionary models}\label{ssec:modelgrid}

\begin{deluxetable}{lcccccccl}
\tablewidth{0pt}
\tablecaption{Description of the model sets considered in this study. The consecutive columns give: `id' -- model set, \fcor{} and \fenv{} -- core and envelope overshooting parameters, \etar{} -- scaling parameter for mass loss, $f_{\rm ST, AM}$ and $f_{\rm ST}$ -- scaling parameters for the diffusion coefficients due to the Spruit–Tayler dynamo for the angular momentum transport and chemical element mixing equations, \fc{} -- a factor that scales the efficiency of chemical element mixing relative to that of angular momentum transport, \fmu{} -- a factor describing the sensitivity of rotationally induced mixing to composition gradients. In the last column, we list the initial rotation rates, $W_0$, adopted in a given model set. \label{tab:models}}
\tablehead{
 \colhead{id} & \colhead{$\fcor$} & \colhead{$\fenv$} & \colhead{$\etar$} & \colhead{$f_{\rm ST, AM}$} & \colhead{$f_{\rm ST}$} & \colhead{$\fc$} & \colhead{$\fmu$} & \colhead{Remarks}
}
\startdata
 \set{O00\_ML4h\_RA}   & 0.00  &  0.00  & 0.4 &  0.0 & 0.0 & 1/30 & 0.05 & $W_0=0,\,0.3,\,0.5,\,0.65$\\ 
 \set{O14\_ML4h\_RA}    & 0.01  &  0.04  & 0.4 &  0.0 & 0.0 & 1/30 & 0.05 & $W_0=0,\,0.3,\,0.5,\,0.65$\\ 
 \set{O24\_ML4h\_RA}    & 0.02  &  0.04  & 0.4 &  0.0 & 0.0 & 1/30 & 0.05 & $W_0=0,\,0.3,\,0.5,\,0.65$\\  
 \set{O14\_ML8h\_RA}    & 0.01  &  0.04  & 0.8 &  0.0 & 0.0 & 1/30 & 0.05 & $W_0=0,\,0.3,\,0.5,\,0.65$\\ 
 \set{O24\_ML8h\_RA}    & 0.02  &  0.04  & 0.8 &  0.0 & 0.0 & 1/30 & 0.05 & $W_0=0,\,0.3,\,0.5,\,0.65$\\  
 \hline
 \set{O14\_ML4h\_RB}    & 0.01  &  0.04  & 0.4 &  1.0 & 0.0 & 1/30 & 0.05 & $W_0=0.5$\\ 
 \set{O14\_ML4h\_RC}    & 0.01  &  0.04  & 0.4 &  1.0 & 1.0 & 1/30 & 0.05 & $W_0=0.5$\\ 
 \set{O14\_ML4h\_RD}    & 0.01  &  0.04  & 0.4 &  0.0 & 0.0 & 1/15 & 0.05 & $W_0=0.5$\\ 
 \set{O14\_ML4h\_RE}    & 0.01  &  0.04  & 0.4 &  0.0 & 0.0 & 1/30 & 0.025 & $W_0=0.5$\\ 
 \enddata
 \tablecomments{In all model sets $f_{\rm ES}$=1, $f_{\rm DSI}$=1, $f_{\rm SSI}$=1, $f_{\rm GSF}$=1, and $f_{\rm SH}$=1.}
 \end{deluxetable}

Our model grid is not as extensive as in \citetalias{Smolec-2026}, as models with rotation are numerically more challenging and time-consuming. The computed model sets are presented in Tab.~\ref{tab:models}. For each model set, we compute the $2-8\MS$ models with a $0.5\MS$ step (13 mass values). Our choice of metallicity is related, on the one hand, to the willingness to present relations appropriate for Cepheids in the MW and in the Magellanic Clouds, and on the other hand, to the willingness to compare the results with those obtained with the Geneva code, presented in \cite{Anderson-2014, Anderson-2016}. Hence, the calculations are performed for $Z=0.014$, $Z=0.006$, and $Z=0.002$. Concerning the overshooting parameters, we fix the envelope overshooting parameter to $\fenv=0.04$, similar to the models discussed in detail in \citetalias{Smolec-2026}. For the MS core overshooting, we consider the models without overshooting $\fcor=0$ (for this model set, also $\fenv=0$) and including overshooting with $\fcor=0.01$ and $\fcor=0.02$; the latter corresponds to the O24 model set, discussed in detail in \citetalias{Smolec-2026}.

For mass loss on the RGB and during core helium burning, we either adopt $\etar=0.4$ or $\etar=0.8$.

In addition to our basic setup of rotation parameters, for a specific model set \set{O14\_ML4h} ($\fcor=0.01$, $\fenv=0.04$, $\etar=0.4$), we also consider four more model sets in which various parameters related to rotation are varied in order to explore their effects on the models. These model sets and their characteristics are included in the bottom section of Tab.~\ref{tab:models}.

Each model set is computed for three different initial rotation rates. We follow \cite{Anderson-2014, Anderson-2016} and assume that $W_0=0.3$ ($\omega_0=0.5$ in the Geneva tracks) represents the average rotation rate of the Cepheid progenitors \citep{Huang-2010}. In addition, we compute the models for higher initial rotation rates of $W_0=0.5$ and $0.65$, the latter corresponding to $\omega_0=0.9$, the second initial rotation rate analyzed in more detail in the Geneva tracks. The models in the bottom section of Tab.~\ref{tab:models} are computed for $W_0=0.5$ only.

\section{Results} \label{sec:results}

\subsection{Overview of evolutionary tracks and online resources}

All computed evolutionary tracks are stored on Zenodo at \url{https://zenodo.org/records/19076324}. An individual track name includes the model set id, initial rotation rate, mass, metal, and helium content values, e.g.,\ \texttt{history.dat\_O12\_ML4h\_RA\_W0.30\_5.5\_0.0140\_0.2695}\ corresponds to a track from the \set{O12\_ML4h\_RA} model set of 5.5\MS, $Z=0.014$, $Y=0.2695$, initialized with $W_0=0.3$. The content (columns) of the evolutionary track files is described in Tab.~\ref{tab:history}. In the online repository, we also include exemplary inlists, one for each model set listed in Tab.~\ref{tab:models} (with $M$/$Z$ fixed to 5\MS{}/$0.002$ and $W_0=0.5$).

\begin{deluxetable}{rll}
\tablewidth{0pt}
\tablecaption{Description of the columns of the evolutionary track files stored in the online repository.\label{tab:history}}
\tablehead{
\colhead{No} & \colhead{Label} & \colhead{Explanation}
}
\startdata
 1 & model\_number & model number \\
 2 & star\_age     & model age in yrs\\
 3 & star\_mass    & model mass (solar units) \\
 4 & log\_Teff     & log effective temperature  \\
 5 & log\_L        & log absolute luminosity (solar units) \\
 6 & log\_R        & log radius (solar units) \\
 7 & log\_g        & log surface gravity (cgs units) \\
 8 & log\_cntr\_P  & log central pressure \\
 9 & log\_cntr\_Rho& log central density\\
10 & log\_cntr\_T  & log central temperature\\
11 & center\_mu    & central mean molecular weight\\ 
12 & center\_h1    & central $^{1}$H  mass fraction \\ 
13 & center\_he4   & central $^{4}$He mass fraction \\ 
14 & center\_c12   & central $^{12}$C mass fraction \\
15 & center\_n14   & central $^{14}$N mass fraction \\
16 & center\_o16   & central $^{16}$O mass fraction \\
17 & surface\_h1   & surface $^{1}$H  mass fraction  \\
18 & surface\_he4  & surface $^{4}$He mass fraction \\
19 & surface\_c12  & surface $^{12}$C mass fraction \\
20 & surface\_n14  & surface $^{14}$N mass fraction \\
21 & surface\_o16  & surface $^{16}$O mass fraction \\
22 & abs\_mag\_V   & absolute magnitude in V band\\
23 & abs\_mag\_I   & absolute magnitude in I band\\
24 & abs\_mag\_J   & absolute magnitude in J band\\
25 & abs\_mag\_H   & absolute magnitude in H band\\
26 & abs\_mag\_K   & absolute magnitude in K band\\
27 & surf\_avg\_v\_rot & rotational velocity at equator (km/s) \\
28 & surf\_avg\_omega\_div\_omega\_crit & $W$, eq.~\eqref{eq:oocritmesa} \\
\enddata
\end{deluxetable}

\subsection{The blue loop evolution}\label{ssec:blueloop_overview}

\begin{figure*}
\includegraphics[width=\linewidth]{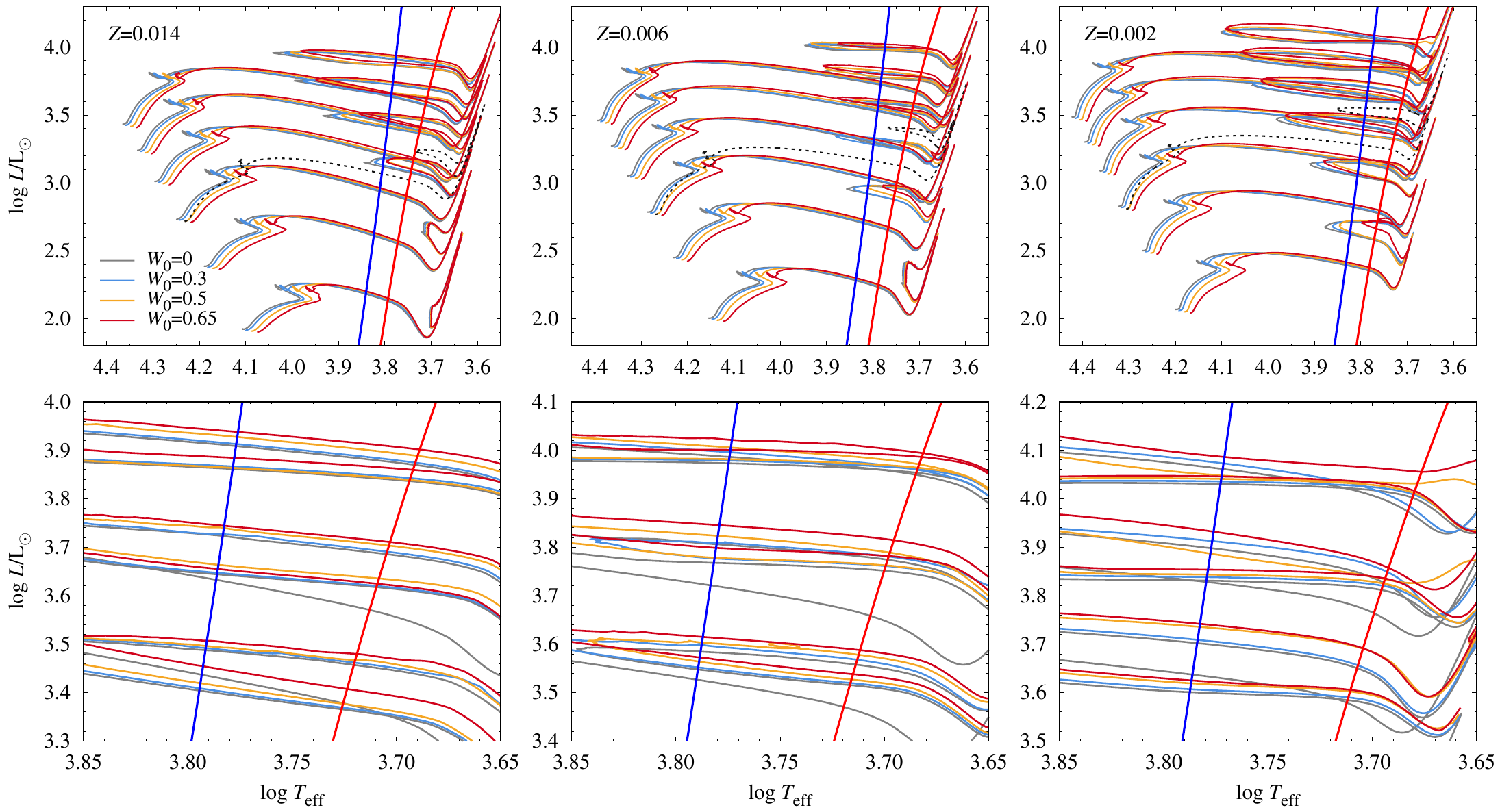}
\caption{Evolutionary tracks for 3, 4, 5, 6, 7, and 8\MS{} models of the \set{O14\_ML4h\_RA} set (with moderate MS core overshooting, $\fcor=0.01$, $\fenv=0.04$, and $\etar=0.4$) for $Z=0.014$ (left), $Z=0.006$ (middle), and $Z=0.002$ (right), and different initial rotation rates coded by color: gray for non-rotating models, then blue for $W_0=0.3$, orange for $W_0=0.5$, and red for $W_0=0.65$. Bottom panels zoom in on the blue loops of 6, 7, and 8\MS{} models. In the top panels we also include 5\MS{}, $W_0=0.5$ tracks computed with a twice as large MS core overshooting parameter ($\fcor=0.02$, black dotted line) for comparison. The edges of the IS are plotted for reference.
\label{fig:hrd_overview}}
\end{figure*}

We discuss the general effects of rotation on evolutionary tracks with the models of the \set{O14\_ML4h\_RA} set, i.e., including moderate MS core overshooting, $\fcor=0.01$. The resulting evolutionary tracks are presented in Fig.~\ref{fig:hrd_overview} for $3-8\MS$ models and the three metallicities we consider in this study. The initial rotation rate is color coded, with gray tracks computed without rotation and then blue, orange, and red colors corresponding to $W_0=0.3$, $0.5$, and $0.65$, respectively. In the top panels we show the entire tracks from the ZAMS until the AGB ascent, while in the bottom panels we show a zoom in to the IS crossings by the blue loops of 6, 7, and 8\MS{} models. In Tab.~\ref{tab:overview}, the properties of some of the tracks are quantified. These are the increase in the MS lifetime with respect to the non-rotating model, $\Delta{\rm Age_{MS}}$, the increase in luminosity at the midpoint of the IS, with respect to the non-rotating model, $\Delta L/\LS$, given separately for each of the three IS crossings, and the average surface rotation velocity at the midpoint of the IS, $\vrot$, for the three IS crossings. The last two columns characterize the luminosity extent of the loop. $L_{23}$ is the luminosity difference between the 3rd and 2nd crossings, evaluated at the midline of the IS. $\Delta L_{23,0}$ is defined as the difference between $L_{23}$ computed for a rotating model and the corresponding value for a non-rotating track.

\begin{deluxetable*}{rrrrrrrrrrrr}
\tablewidth{0pt}
\tablecaption{Selected properties of evolutionary tracks for \set{OV14\_ML4h\_RA} models ($\fcor=0.01$, $\fenv=0.04$, $\etar=0.4$) of mass, metallicity, and initial rotation rate given in the first three columns. The properties are the increase in MS lifetime with respect to the non-rotating model of the same $M$/$Z$, $\Delta{\rm Age_{MS}}$ (in per cent), followed by the increase in luminosity with respect to the non-rotating track, $\Delta L/\LS$, and the average surface rotational velocity, $\vrot$, given separately for the three crossings of the IS and evaluated at the midline of the IS. In the last two columns, $L_{23}$ and $\Delta L_{23,0}$ characterize the thickness of the loop, i.e., they correspond to the luminosity difference between the 2nd and 3rd crossings, evaluated at the IS midline, for the rotating model ($L_{23}$), and the difference with respect to the non-rotating model ($\Delta L_{23,0}$). No data means that the loop does not reach the midline of the IS for a given model (either for the model including rotation or for the non-rotating model, which is necessary, e.g., to compute $\Delta L/\LS$ values).
\label{tab:overview}}
\tablehead{
  &     &  &  & \multicolumn{2}{c}{1st crossing}  & \multicolumn{2}{c}{2nd crossing}   & \multicolumn{2}{c}{3rd crossing} & \\
\colhead{$M$} & \colhead{$Z$} & \colhead{$W_0$} & \colhead{$\Delta$\,Age$_{\rm MS}$} & \colhead{$\Delta L/\LS$} & \colhead{$v_{\rm rot}$} &  \colhead{$\Delta L/\LS$} & \colhead{$v_{\rm rot}$} &  \colhead{$\Delta L/\LS$} & \colhead{$v_{\rm rot}$} & \colhead{$L_{23}$} & \colhead{$\Delta L_{23,0}$}}
\startdata
4.0 & 0.014 &  0.3 &  1.5 & 0.006 & 47.5 &      -   &    - &        - &    - &     - & - \\
4.0 & 0.014 &  0.5 &  3.5 & 0.015 & 81.0 &        - &    - &        - &    - &     - & - \\
4.0 & 0.014 & 0.65 &  6.0 & 0.023 & 94.4 &        - &    - &        - &    - &     - & - \\
4.0 & 0.006 &  0.3 &  1.7 & 0.007 & 40.9 &  $0.002$ & 46.1 &  $0.005$ & 44.4 & 0.076 &  $0.004$ \\
4.0 & 0.006 &  0.5 &  4.1 & 0.018 & 70.7 &  $0.032$ & 85.7 &  $0.005$ & 75.6 & 0.046 &  $-0.026$ \\
4.0 & 0.006 & 0.65 &  7.0 & 0.026 & 83.3 &        - &    - &        - &    - &     - &         - \\
4.0 & 0.002 &  0.3 &  1.9 & 0.007 & 35.0 &  $0.010$ & 44.0 &  $0.006$ & 36.6 & 0.100 & $-0.004$ \\
4.0 & 0.002 &  0.5 &  5.1 & 0.020 & 61.0 &  $0.021$ & 77.2 &  $0.019$ & 54.0 & 0.101 & $-0.003$ \\
4.0 & 0.002 & 0.65 &  8.4 & 0.023 & 72.7 &  $0.011$ & 95.4 &  $0.033$ & 57.6 & 0.127 & $0.023$ \\
5.0 & 0.014 &  0.3 &  1.6 & 0.008 & 36.2 &  $0.010$ & 38.6 &  $0.004$ & 39.7 & 0.073 & $-0.006$ \\
5.0 & 0.014 &  0.5 &  3.8 & 0.017 & 62.1 &  $0.036$ & 72.3 & $-0.001$ & 68.6 & 0.042 & $-0.037$ \\
5.0 & 0.014 & 0.65 &  6.6 & 0.026 & 71.6 &  $0.031$ & 85.4 &  $0.007$ & 77.7 & 0.055 & $-0.024$ \\
5.0 & 0.006 &  0.3 &  1.7 & 0.008 & 30.9 & $-0.004$ & 38.8 &  $0.011$ & 34.8 & 0.057 & $0.016$ \\
5.0 & 0.006 &  0.5 &  4.7 & 0.023 & 54.0 &        - &    - &        - &    - &     - &    - \\
5.0 & 0.006 & 0.65 &  7.5 & 0.027 & 63.1 &        - &    - &        - &    - &     - &    - \\
5.0 & 0.002 &  0.3 &  1.9 & 0.009 & 26.5 &  $0.008$ & 34.4 &  $0.011$ & 28.4 & 0.113 & $0.003$ \\
5.0 & 0.002 &  0.5 &  5.2 & 0.017 & 46.3 &  $0.004$ & 61.4 &  $0.027$ & 42.4 & 0.134 & $0.024$ \\
5.0 & 0.002 & 0.65 & 10.2 & 0.041 & 54.8 &  $0.040$ & 77.8 &  $0.038$ & 44.6 & 0.108 & $-0.002$ \\
7.0 & 0.014 &  0.3 &  1.6 & 0.008 & 22.4 &  $0.003$ & 28.5 &  $0.005$ & 26.7 & 0.076 & $0.003$ \\
7.0 & 0.014 &  0.5 &  4.6 & 0.019 & 38.2 &  $0.018$ & 51.2 &  $0.016$ & 41.9 & 0.072 & $-0.001$ \\
7.0 & 0.014 & 0.65 &  7.9 & 0.035 & 44.9 &  $0.009$ & 59.2 &  $0.023$ & 43.9 & 0.088 & $0.015$ \\
7.0 & 0.006 &  0.3 &  1.7 & 0.009 & 19.8 &  $0.009$ & 29.2 &  $0.004$ & 24.8 & 0.027 & $-0.004$ \\
7.0 & 0.006 &  0.5 &  5.1 & 0.021 & 34.0 &  $0.010$ & 52.9 &  $0.016$ & 36.2 & 0.037 & $0.006$ \\
7.0 & 0.006 & 0.65 &  9.4 & 0.041 & 40.0 &  $0.029$ & 66.9 &  $0.040$ & 38.0 & 0.041 & $0.010$ \\
7.0 & 0.002 &  0.3 &  2.0 & 0.012 & 17.3 &  $0.008$ & 23.4 &  $0.018$ & 20.1 & 0.058 & $0.011$ \\
7.0 & 0.002 &  0.5 &  6.7 & 0.021 & 30.0 &  $0.013$ & 42.2 & $-0.012$ & 30.1 & 0.022 & $-0.025$ \\
7.0 & 0.002 & 0.65 & 12.0 & 0.040 & 35.1 &  $0.024$ & 51.1 &  $0.030$ & 31.0 & 0.054 & $0.007$ \\
8.0 & 0.014 &  0.3 &  1.7 & 0.008 & 18.1 &  $0.006$ & 25.7 &  $0.004$ & 21.3 & 0.039 & $-0.001$ \\
8.0 & 0.014 &  0.5 &  4.7 & 0.023 & 30.6 &  $0.002$ & 42.9 &  $0.019$ & 31.6 & 0.057 & $0.017$ \\
8.0 & 0.014 & 0.65 &  7.7 & 0.028 & 36.1 &  $0.023$ & 53.5 &  $0.030$ & 34.3 & 0.048 & $0.008$ \\
8.0 & 0.006 &  0.3 &  1.8 & 0.008 & 16.3 &  $0.006$ & 25.0 &  $0.010$ & 19.9 & 0.009 & $0.004$ \\
8.0 & 0.006 &  0.5 &  5.3 & 0.016 & 27.7 &  $0.010$ & 44.1 &  $0.020$ & 29.4 & 0.016 & $0.011$ \\
8.0 & 0.006 & 0.65 &  8.8 & 0.025 & 33.0 &  $0.030$ & 59.5 &  $0.042$ & 33.0 & 0.017 & $0.012$ \\
8.0 & 0.002 &  0.3 &  1.9 & 0.006 & 14.6 &  $0.006$ & 20.5 &  $0.012$ & 17.6 & 0.014 & $0.006$ \\
8.0 & 0.002 &  0.5 &  6.7 & 0.018 & 25.4 &  $0.011$ & 35.3 &  $0.005$ & 25.7 & 0.002 & $-0.006$ \\
8.0 & 0.002 & 0.65 & 11.7 & 0.076 & 29.5 &  $0.015$ & 41.1 &  $0.038$ & 25.4 & 0.031 & $0.023$ \\
\enddata
\end{deluxetable*}

In Fig.~\ref{fig:hrd_overview}, we observe the following. Independent of metallicity, with increasing initial rotation rate, the MS tracks shift towards lower effective temperatures. This is mainly the effect of the centrifugal force; the envelope slightly expands, leading to a lower effective temperature. The tracks are shifted nearly horizontally, i.e., we do not observe any significant shift in the MS luminosity. The MS hook and TAMS stay at nearly the same luminosity level, independent of the initial rotation rate. Considering the MS evolution time, it slightly increases with increasing initial rotation rate. This is quantified in the fourth column of Tab.~\ref{tab:overview} for models of 4, 5, 7, and 8\MS{}, and for different metallicities. For $W_0=0.3$ the increase in the MS lifetime is typically by up to 2\%, then for $W_0=0.5$ by about 5\%, and finally for the largest initial rotation rate, $W_0=0.65$, by 6 to 12\%. At a given initial rotation rate, the increase is larger for lower metallicities and larger masses. The increased MS lifetime is due to rotational mixing at the border of the hydrogen-burning convective core. While slightly extending the MS lifetime, this mixing is not efficient enough to significantly increase the MS luminosity.

After the MS, the evolutionary tracks corresponding to different initial rotation rates follow very similar paths, with slight differences being apparent only once core helium burning commences.

In the context of classical Cepheids, particularly the mass discrepancy problem, it is crucial whether the blue loop luminosity increases with increasing rotation rate. As can be seen more clearly in the close-ups in the lower panels of Fig.~\ref{fig:hrd_overview}, this is indeed the case, but the increase in brightness is very small. For the individual IS crossings, the luminosity increase with respect to the non-rotating tracks, $\Delta L/\LS$, is quantified in Tab.~\ref{tab:overview}. The values given correspond to the intersections of the tracks with the IS midline. In general, we observe an increase in luminosity, relative to the non-rotating tracks, with increasing rotation rate, but the highest recorded increase is only about 0.04\,dex. This is much smaller than, e.g., the increase in luminosity resulting from increasing the MS core overshooting from $\fcor=0.01$ to $\fcor=0.02$. To illustrate this point, in the top panels of Fig.~\ref{fig:hrd_overview}, we have also included evolutionary tracks for 5\MS{}, $W_0=0.5$ models with \fcor=0.02 (dotted black lines). Such a change leads to about a $0.1-0.15$\,dex luminosity increase (depending on the crossing number and metallicity) during the blue loop phase.

Of particular interest is the luminosity extent of the loop, characterized by the $L_{23}$ and $\Delta L_{23,0}$ parameters (last columns of Tab.~\ref{tab:overview}). General trends should be considered rather than individual values, which may appear peculiar. The $L_{23}$ parameter is evaluated from the crossings of the track with the midline of the IS. When the loop turns within the IS, close to its midline, $L_{23}$ is small, as the loop is thin near the turning point. In such cases, even small differences in the temperature extent between the rotating and non-rotating tracks may lead to large differences in $L_{23}$ (and consequently in $\Delta L_{23,0}$), as the midline may intersect one loop near its tip and the other where it is already significantly thicker. We observe that the loops are, in general, relatively thin; the loop extent may slightly exceed 0.1\,dex only for lower-mass models. For higher masses ($M>5\MS$), it typically remains within 0.02--0.06\,dex. Compared to the non-rotating tracks, we do not observe any significant differences. Only for higher-mass, fast-rotating models ($M>5\MS$, $W_0\geq0.5$) is a thicker loop more frequently recorded for the rotating tracks (positive $\Delta L_{23,0}$), but the difference rarely exceeds 0.01\,dex.

The scenario for models without overshooting (\set{O00\_ML4h\_RA}) and with higher MS core overshooting (\fcor=0.02, \set{O24\_ML4h\_RA}) is qualitatively the same as discussed above. We do not observe any significant increase in the blue loop luminosity due to rotation in these models.

The recorded increase in luminosity is definitely too small to offer a solution to the mass discrepancy problem in terms of rotation (i.e., without simultaneously requiring significant MS core overshooting), contrary to the results obtained with the Geneva code, as we discuss in more detail by analyzing the \ML{} relation in Sect.~\ref{ssec:ml} and discuss in Sect.~\ref{ssec:compgeneva}.

Another property of interest is the surface rotation velocity, $v_{\rm rot}$, during the blue loop phase. These velocities are, in general, quite large, as quantified in Tab.~\ref{tab:overview}, at the midpoints of the IS, separately for each crossing. Even for the lowest initial rotation rate, $W_0=0.3$, the surface rotation velocities at the 2nd crossing vary from about 45\,km/s (for 4\MS) to about 20--25\,km/s (for 8\MS). Only during the 3rd crossing and for 8\MS{}, does the rotation velocity drop below 20\,km/s. For larger initial rotation rates, the predicted rotation velocities during the blue loop phase are even higher. These values are well above those expected for classical Cepheids, for which \cite{Nardetto-2006} gives $V_{\rm rot}\sin i = (-11.5\pm0.9)\log P + (19.8\pm1.0)$ [km/s]. Assuming that F-mode pulsation periods for the majority of Cepheids are between 1 and 10\,d, typical surface rotation velocities should be from around 10\,km/s (long pulsation periods) to about 25\,km/s (short pulsation periods). An average equatorial rotational velocity for a sample of 97 MW Cepheids given by \cite{Anderson-PhD} is 12.3\,km/s.

\begin{figure}
\includegraphics[width=\linewidth]{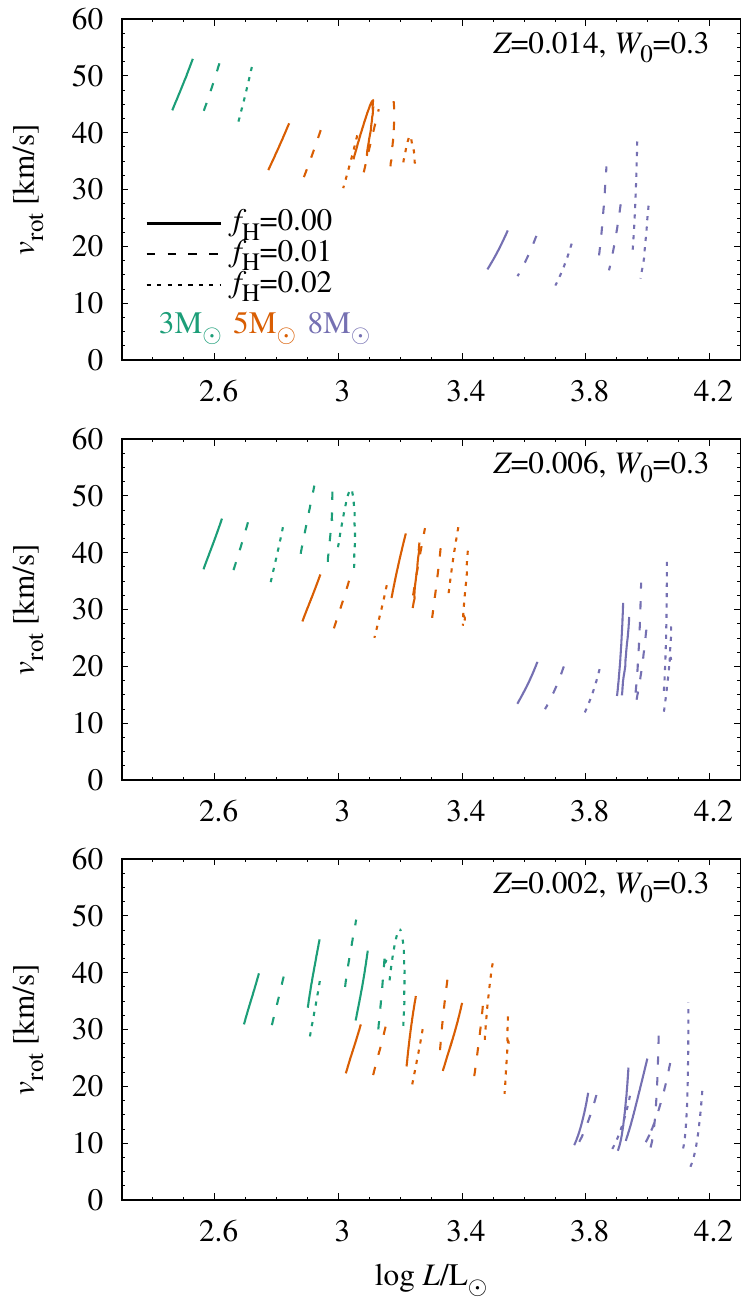}
\caption{Average surface rotational velocities, \vrot{}, as a function of luminosity during crossings of the IS for 3, 5, and 8\MS{} models of different metallicities, $Z=0.014$ (top), $Z=0.006$ (middle), and $Z=0.002$ (bottom). Solid, dashed, and dotted lines correspond to models without MS core overshooting, and with $\fcor=0.01$, and $\fcor=0.02$, respectively. In all models, the initial ZAMS rotation rate is $W_0=0.3$.
\label{fig:vrot_overview}}
\end{figure}

To investigate rotational velocities further, in Fig.~\ref{fig:vrot_overview} we show the evolution of \vrot{} during the three crossings of the IS, for 3, 5, and 8\MS{} models (different colors) of $Z=0.014$ (top), $Z=0.006$ (middle), and $Z=0.002$ (bottom). In all models, the lowest initial rotation rate is assumed, $W_0=0.3$. Each line segment corresponds to the evolution of \vrot{} during a full crossing of the IS. The 1st, 2nd, and 3rd crossings may be distinguished by increasing luminosity. Since the evolutionary track is nearly horizontal during the IS crossing, the evolution of \vrot{} is mostly due to the change in radius. During the 1st and 3rd crossings, the star expands, and thus the surface spins down. During the 2nd crossing, the star contracts and spins up. In Fig.~\ref{fig:vrot_overview} we include models without MS core overshooting (solid lines), as well as with $\fcor=0.01$ (dashed lines), and $\fcor=0.02$ (dotted lines).

We observe that for models of the same initial surface rotation rate, as displayed in Fig.~\ref{fig:vrot_overview}, the surface rotation velocity depends primarily on mass and metallicity, decreasing with increasing mass and decreasing metallicity. It depends little on the crossing number. Surface rotation velocities are similar during the 2nd and 3rd crossings and are, on average, slightly larger than during the 1st crossing. Within each crossing, the rotation velocity changes by about 10 to 20\,km/s.

Surface rotation velocity depends very little on the MS overshooting. The ranges recorded for $\fcor=0.0$, $\fcor=0.01$, and $\fcor=0.02$ models are very similar, the differences being of the order of a few km/s.

As shown in Fig.~\ref{fig:vrot_overview}, the predicted surface rotation velocities are significantly higher than those expected for classical Cepheids. The only exception is the most massive 8\MS{} models, in which, for the lowest metallicity, the rotation velocities drop below 10,km/s. The very modest rotation-induced increase in blue loop luminosities and high surface rotation velocities constitute the two main issues associated with the \MESA{} implementation of rotation in the context of classical Cepheid models.

The modest impact of rotation on evolutionary tracks likewise implies a small effect on evolutionary period change rates (PCRs). This is indeed the case. For $W_0=0.3$, a typical value for Cepheid progenitors, the predicted PCRs differ only marginally from those computed for non-rotating models. In the Appendix (Sect.~\ref{secapp:pcr}), we explicitly compare PCRs derived from models without rotation and with rotation (Fig.~\ref{figapp:pcr}). We also provide PCR values for all model sets listed in the upper section of Tab.~\ref{tab:models} in tabular form. The table is analogous to Tab.~6 of \citetalias{Smolec-2026} and is available in its entirety in the electronic edition of the Journal and on Zenodo. We further refer the reader to \cite{EspinozaArancibia-2022}, who presented a dedicated study of PCRs in classical Cepheid models including rotation, computed with \MESA.

\subsection{The impact of free parameters}\label{ssec:freep}

In the previous section, we have investigated how mass, metallicity, MS core overshooting, \fcor, and initial rotation rate, $W_0$, impact the blue loops and their properties. The rotation model also has some free parameters that control the diffusive mixing processes and their efficiency. In addition, mass loss affects the rate of angular momentum transfer. Now we investigate how the associated parameters affect evolutionary tracks and the surface rotation velocity at the blue loop phase.

So far, we have neglected the transport of angular momentum and chemical element mixing due to internal magnetic fields. We study these effects in two model sequences. In the first, \set{O14\_ML4h\_RB} (see Tab.~\ref{tab:models}), we include magnetic angular momentum transport via the Spruit–Tayler (ST) dynamo ($f_{\rm ST, AM}=1$). In the second, \set{O14\_ML4h\_RC}, we also include chemical element mixing due to the same mechanism ($f_{\rm ST, AM}=1$, $f_{\rm ST}=1$).

The overall efficiency of chemical element mixing, with respect to angular momentum transport, is scaled with the $\fc$ parameter. Another free parameter, \fmu, describes the sensitivity of rotationally induced mixing to composition gradients (Sect.~\ref{ssec:methods_rotation}). For these two parameters we use $\fc=1/30$ and $\fmu=0.05$ by default. The values of these parameters were calibrated to reproduce surface abundances of elements such as lithium or nitrogen \citep[e.g.,][]{Heger-2000}. Calibration of lithium abundance in the Sun results in $\fc=1/30$ \citep[see also][]{ChaboyerZahn-1992,Heger-2000}. \cite{Heger-2000} adopted $\fmu=0.05$ as best reproducing helium and nitrogen enrichment in massive stars. In the sequence \set{OV14\_ML4h\_RD}, we increase the efficiency of chemical element mixing by setting $\fc$ to twice the default value, $\fc=1/15$, as more efficient mixing should increase the luminosity during MS evolution as well as in the following evolutionary phases. In the sequence \set{OV14\_ML4h\_RE}, we decrease \fmu{} to $0.025$, as it should ensure more efficient mixing even for stabilizing composition gradients. Our aim is to examine whether factor-of-two variations in these parameters lead to significant changes in the blue-loop luminosity and the surface rotational velocity. The adopted values may, in fact, conflict with other observational constraints. For example, in order to reproduce the main trend of the observed nitrogen surface abundances in nitrogen-enriched, rapidly rotating B-type stars in the LMC as a function of projected rotational velocity, \citet{Brott-2011} adopted a smaller value of $\fc=0.0228$.

\begin{figure}
\includegraphics[width=\linewidth]{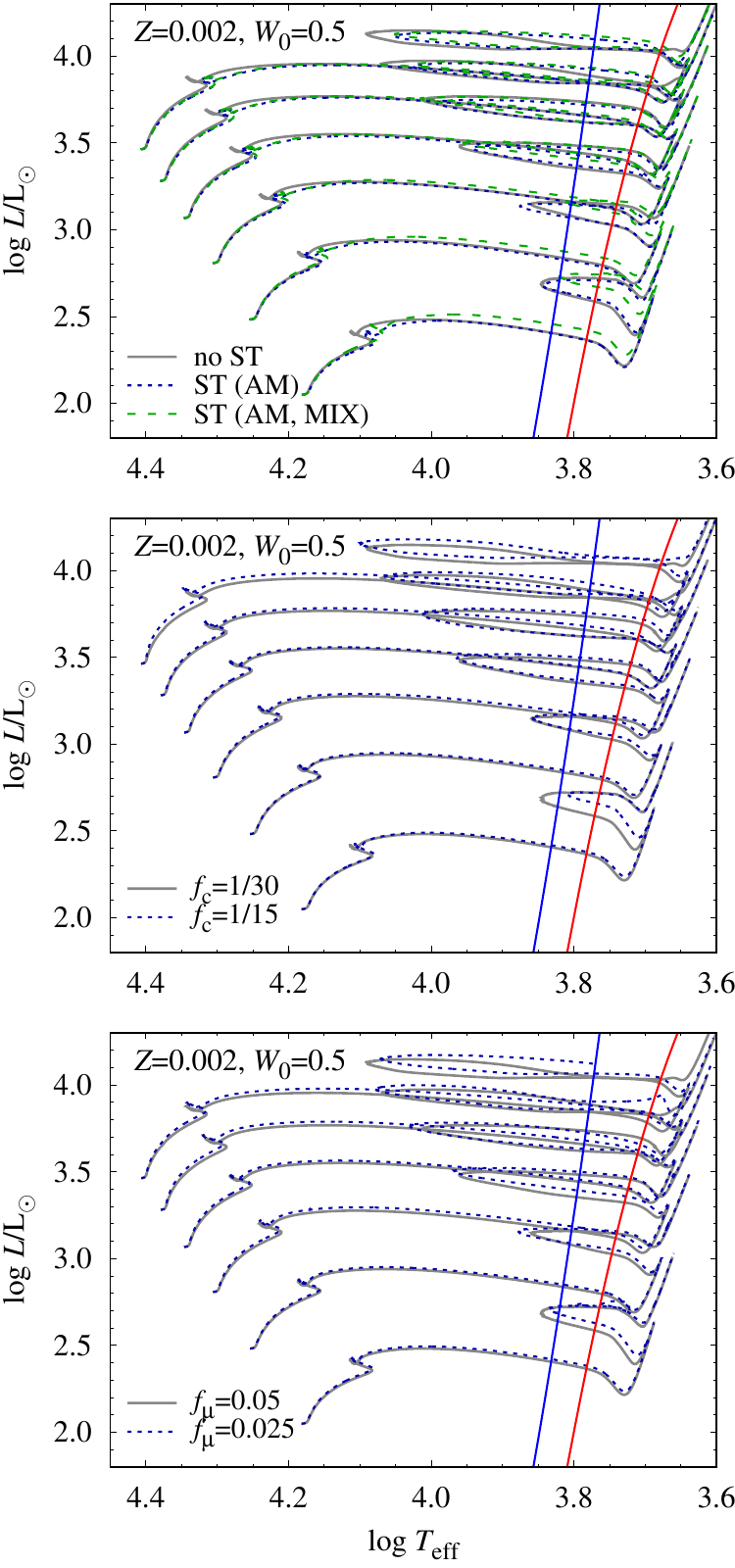}
\caption{HRDs for $3-8\MS$, $Z=0.002$ models, initialized with a rotation rate of $W_0=0.5$, with different rotation-related controls aimed at increasing blue loop luminosity. In the top panel, the effects of inclusion of magnetic effects (ST dynamo) in angular momentum (dotted blue tracks) and simultaneously in chemical element mixing (dashed green tracks) are illustrated. In the middle panel, we compare tracks computed with the default value of $\fc=1/30$ and a twice as large value, $\fc=1/15$. In the bottom panel, tracks with $\fmu=0.05$ (the default) and a factor of two smaller value, $\fmu=0.025$, are compared.
\label{fig:hrd_params}}
\end{figure}

In Fig.~\ref{fig:hrd_params}, we show how the discussed changes affect the evolutionary tracks. Since the observed trends are similar for all metallicities, we show only the results for $Z=0.002$, as for this metallicity the blue loops are well developed for all masses. The top panel shows the effect of including the ST contribution to angular momentum transport (dotted blue lines) and chemical element mixing (dashed green lines) on the evolutionary tracks. Inclusion of ST in angular momentum transport only very little affects the evolutionary tracks. The luminosity levels of the blue loops either do not change significantly or are slightly decreased. The exception are the two most massive models of 7 and 8\MS{}, for which the luminosity during the 3rd crossing is significantly increased, by about 0.05\,dex. This is a consequence of the change in the shape of the loop. Models without ST have a wide loop (a significant separation between the 2nd and 3rd crossings) outside the blue edge of the IS. Inside the IS, the separation decreases significantly; for the 8\MS{} model, the luminosity at the 2nd and 3rd crossings is almost the same. Inclusion of the ST contribution in angular momentum transport (as well as simultaneously in the mixing of chemical elements) causes the loop to remain wide also inside the IS, which results in a relative increase in luminosity at the 3rd crossing. When ST is included also in the chemical element mixing, a luminosity increase is observed for all masses, but most significantly for the 2nd crossing in low-mass models ($3-5\MS$, by up to 0.07\,dex with respect to models without ST) and for the 3rd crossing for the most massive models (7, 8\MS), for the reason outlined above. For the 6\MS{} model, the luminosity at both crossings is nearly unaffected.

\begin{figure}
\includegraphics[width=\linewidth]{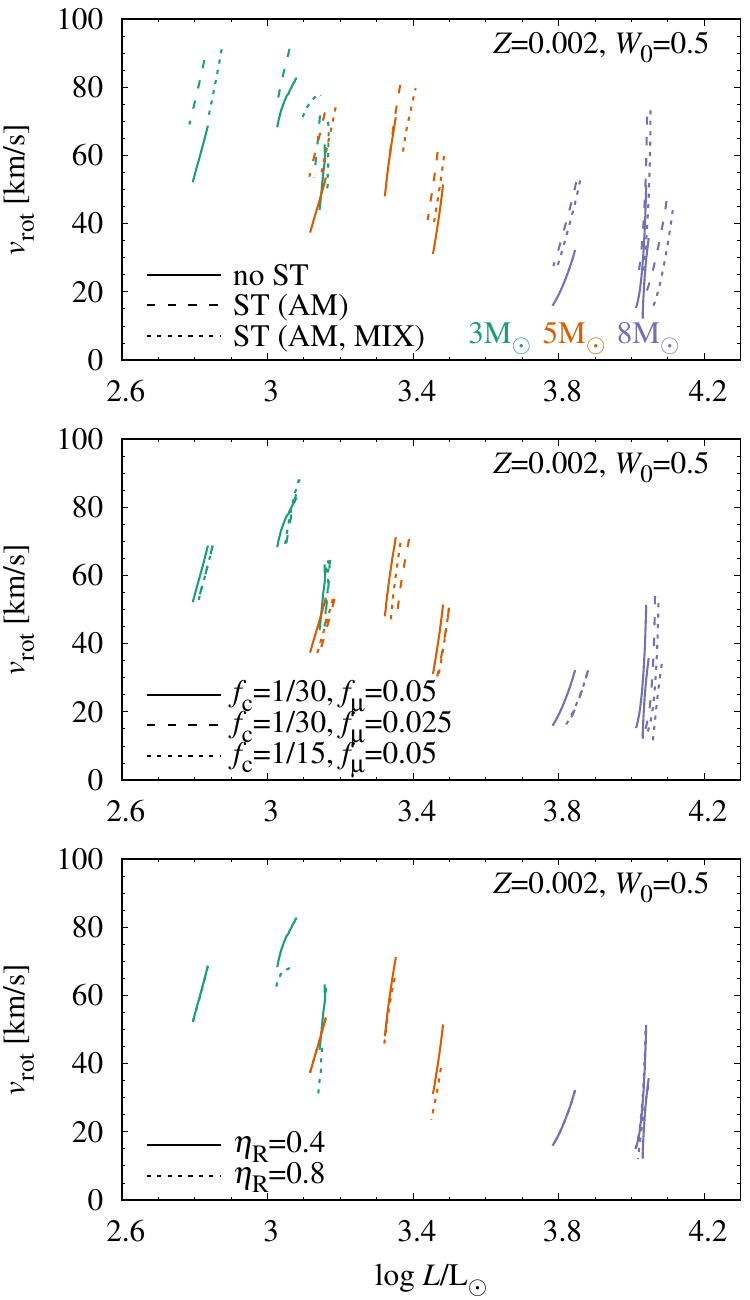}
\caption{Average surface rotational velocities as a function of luminosity during crossings of the IS for 3, 5, and 8\MS{} models of $Z=0.002$, initialized with a rotation rate of $W_0=0.5$. In the top panel, models without magnetic effects and models including magnetic effects are compared. In the middle panel, models with different values of parameters affecting chemical element mixing, $\fc$ and $\fmu$, are compared. In the bottom panel, models with different efficiencies of mass loss (\etar=0.4 vs.\ \etar=0.8) are compared.
\label{fig:vrot_params}}
\end{figure}

While in general the inclusion of the ST contribution slightly increases the blue loop luminosity, at the same time it significantly increases the surface rotation velocities. This is illustrated in the top panel of Fig.~\ref{fig:vrot_params}. Depending on the mass and crossing number, the increase is by about 10 to 20\,km/s. This increase in surface rotation velocity is expected. After the MS, the core spins up. The ST mechanism transfers angular momentum from the core towards the surface, slowing down the core but spinning up the envelope. While such a change is desired to reproduce asteroseismic determinations of core rotation rates in low-mass red giants \citep[for which models still predict too fast core rotation, even when ST is included, see][]{Cantiello-2014,Fuller-2019}, for the blue loop phase the tension with observations increases even more.

In the middle panel of Fig.~\ref{fig:hrd_params}, we show the effects of a factor-of-two increase of \fc{} on the evolutionary tracks. As expected, we observe an increase in luminosity, starting from the MS, due to increased chemical element mixing efficiency. The increase is very modest, however. At the blue loop phase it depends on the mass and crossing number, but on average it amounts to only $0.02-0.03$\,dex. The largest increase of 0.07\,dex is recorded for the 2nd crossing of the 3\MS{} model.

For decreased \fmu{} (bottom panel of Fig.~\ref{fig:hrd_params}), we observe qualitatively the same changes as when \fc{} was increased, an overall slight increase in luminosity, of the order of $0.02-0.04$\,dex.

The changes in both \fc{} and \fmu{} barely affect the surface rotation velocity during the blue loop phase, which is illustrated in the middle panel of Fig.~\ref{fig:vrot_params}. The rotation velocities are essentially the same as in the models with the default values of \fc{} and \fmu.

Considering mass loss, it very little affects the evolutionary tracks; the effects are similar to those described for non-rotating models in \citetalias{Smolec-2026}, and we do not visualize them (the tracks would largely overlap). For the surface rotation velocity, we expect its decrease, as angular momentum is more efficiently removed from the surface layers with an increased mass-loss rate. This is illustrated in the bottom panel of Fig.~\ref{fig:vrot_params}. This decrease is very small, however, amounting to a few km/s for low-mass models, and is, in fact, barely noticeable for the most massive models.

\subsection{Data for evolutionary and pulsation relations\label{ssec:datarel}}

In the following sections, we analyze the $\ML$, $\PL$, $\PR$, and $\PA$ relations. These are evaluated either at the edges of the IS, blue and red, or along its midline, and may be derived separately for each model set, crossing number, metallicity, and initial rotation rate. The data used to construct the relations are compiled in a table available in the electronic edition of the Journal and on Zenodo\footnote{\url{https://zenodo.org/records/19076324}}. The content of this table is described in Tab.~\ref{tab:datatab_content}. The dataset includes all model sets listed in the top section of Tab.~\ref{tab:models}. The tabulated quantities were obtained from the evolutionary tracks by determining their intersections with the IS boundaries and midline. For details on the bolometric magnitudes, see \citetalias{Smolec-2026}. Linear pulsation periods of the fundamental mode were computed with \RSP{}.

While below we provide analytical fits to the discussed relations, their scope is limited to selected model sets and initial rotation rates. For comparison with observations and for follow-up studies, we recommend using the tabular data directly.

Since our model survey is limited to three metallicity values, unlike in \citetalias{Smolec-2026}, we will not derive metallicity-dependent relations; the metallicity dependence will be discussed by directly comparing relations for different $Z$. For all the relations we will fit the following linear relation in logarithmic space,
\begin{equation}
\log y = a\left(\log x - \log x_0\right)+b\,. \label{eq:simplefit}
\end{equation}
For centering, $x_0$, we will adopt the same values as we used in \citetalias{Smolec-2026}, to allow a more direct comparison with the relations presented therein.

\begin{deluxetable}{rll}
\tablewidth{0pt}
\tablecaption{Content of the table available in the electronic edition of the Journal and online on Zenodo, with data for the evolutionary and pulsation relations.\label{tab:datatab_content}}
\tablehead{
\colhead{Col.} & \colhead{Label} & \colhead{Explanation}}
\startdata
1 & edge & IS identifier (b/r/m -- blue/red/midline) \\
2 & cross. & crossing number \\
3 & set & model set identifier, see Tab.~\ref{tab:models} \\
4 & $W_0$ & initial surface angular velocity \\
5 & $M/\MS$ & model mass \\
6 & $Z$ & metal mass fraction \\
7 & $X$ & hydrogen mass fraction \\
8 & log age & logarithm of age \\
9 & $\log\Teff$ & logarithm of effective temperature \\
10 & $\log L/\LS$ & logarithm of luminosity \\
11 & $\log R/\RS$ & logarithm of radius \\
12 & $Y_c$ & central helium mass fraction\\
13 & $P_0$ & F mode pulsation period \\
   &       & (`x' if \RSP{} model didn't converge) \\
14 & $P_1$ & 1O pulsation period \\
   &       & (`x' if \RSP{} model didn't converge) \\
15 & $V$ & absolute magnitude in $V$ band \\
16 & $I$ & absolute magnitude in $I$ band \\
17 & $J$ & absolute magnitude in $J$ band \\
18 & $H$ & absolute magnitude in $H$ band \\
19 & $K$ & absolute magnitude in $K$ band \\
20 & $\vrot$ & average surface equatorial velocity\\
   &  & (in km/s) \\
21 & $W$ & eq.~\eqref{eq:oocritmesa}\\
\enddata
\end{deluxetable}

\subsection{Mass-Luminosity relation}\label{ssec:ml}

\begin{figure*}
\includegraphics[width=\linewidth]{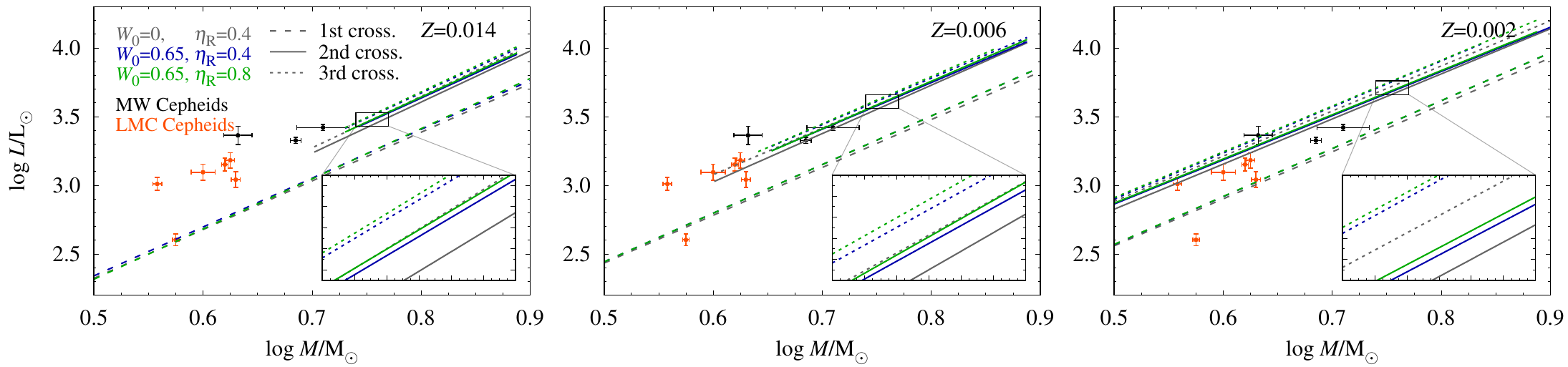}
\caption{$\ML$ relations for \set{O24\_ML4h\_RA} models ($\fcor=0.02$, $\fenv=0.04$, $\etar=0.4$) with $W_0=0$ (gray lines) and the highest initial rotation rate of $W_0=0.65$ (blue lines), and for models of the \set{O24\_ML8h\_RA} set (with twice as large $\etar=0.8$) and $W_0=0.65$, for MW, LMC, and SMC metallicities ($Z=0.014$, $0.006$, and $0.002$ in the left, middle, and right panels, respectively). The theoretical relations are confronted with determinations from \cite{Gallenne-2018-V1334Cyg, Gallenne-2025-SuCyg, Pilecki-2018, Evans-2024-Polaris}.
\label{fig:ml}}
\end{figure*}

We derive the \ML{} relation for the midline of the IS, separately for each crossing, metallicity, and initial rotation rate, $W_0$. The coefficients of analytical fits to eq.~\eqref{eq:simplefit}, for selected model sets, are collected in Tab.~\ref{tab:ml}. In Fig.~\ref{fig:ml} we display the \ML{} relations for models of the \set{O24\_ML4h\_RA} set ($\fcor=0.02$, $\fenv=0.04$, $\etar=0.4$) without rotation, $W_0=0$ (gray lines), and with the highest considered initial rotation rate, $W_0=0.65$ (blue lines). We also show the results for the \set{O24\_ML8h\_RA} set, i.e., with a twice as high mass-loss rate, $\etar=0.8$, also with $W_0=0.65$ (green lines). Separate panels display the \ML{} relations for the three metallicities considered.

The first striking, but expected observation is that for a given crossing of the IS, all these relations, whether without rotation or with the largest initial rotation rate, nearly overlap. While for the first crossing the shift between the $W_0=0$ and $W_0=0.65$ relations can barely be noted in the figure (in particular for larger masses), differences for the loop relations become visible only in the zoomed-in inset panels. Rotation does increase the luminosity level of the blue loops, as we presented in Fig.~\ref{fig:hrd_overview} and discussed in Sect.~\ref{ssec:blueloop_overview}, however the increase in luminosity is very modest. Now we have quantified this increase with the \ML{} relations. Note that we have plotted the relations for the largest initial rotation rate. For $W_0=0.3$ (which corresponds to the typical rotation rate of Cepheid progenitors), or $W_0=0.5$, the differences would be even smaller. In the zoomed-in insets we can observe that the shifts between the relations corresponding to $W_0=0$ (no rotation) and to $W_0=0.65$ increase with decreasing metallicity, which is consistent with the observations made in Sect.~\ref{ssec:blueloop_overview}. Still, for $Z=0.002$, the separation between the two relations for the 2nd crossing is, on average, 0.023\,dex, and for the 3rd crossing 0.031\,dex.

When stronger mass loss is included in the model ($\etar=0.8$; compare the relations plotted with blue and green lines in Fig.~\ref{fig:ml}), the relations, as expected, slightly shift to the left, i.e., at a given luminosity level, the corresponding mass is lower for the model with more efficient mass loss. The shift, however, is significantly smaller (by a factor of $\approx$4) than that resulting from the inclusion of rotation.

In Fig.~\ref{fig:ml}, we also show mass and luminosity determinations for a few Cepheids from \cite{Gallenne-2018-V1334Cyg, Gallenne-2025-SuCyg, Pilecki-2018, Evans-2024-Polaris}. As rotation does not affect the \ML{} relation in a significant way, the comparison with the observed data leads to the same conclusion as in \citetalias{Smolec-2026}. The inclusion of significant MS core overshooting is necessary to match the observed \ML{} relation. Nevertheless, at least some of the observed Cepheids appear too bright compared with the model predictions for a given mass. A lower metallicity leads to higher luminosities. The rotation-induced increase in blue-loop luminosity is very modest and therefore does not provide an alternative to MS core overshooting.

These results are in contrast with those of \cite{Anderson-2014, Anderson-2016}. For Geneva models, rotation does impact the \ML{} relation significantly. A detailed comparison will be presented in Sect.~\ref{ssec:compgeneva}.

\begin{deluxetable*}{lllrrrrrrrrrr}
\tablecaption{Coefficients of the $\ML$ relations as given by eq.~\eqref{eq:simplefit}, computed along the midline of the IS. The first three columns identify the data set, metallicity, and initial rotation rate. Then coefficients of the linear relation, eq.~\eqref{eq:simplefit}, and rms follow, separately for the three crossings of the IS. In the last column, we provide the parameter range over which the 2nd and 3rd crossing relations were derived.\label{tab:ml}}
\tablehead{
 & & & \multicolumn{3}{c}{1st cross.} & \multicolumn{3}{c}{2nd cross.} & \multicolumn{3}{c}{3rd cross.} & \\
\colhead{data} & \colhead{$Z$} & \colhead{$W_0$} & \colhead{$a$} & \colhead{$b$} & \colhead{rms} & \colhead{$a$} & \colhead{$b$} & \colhead{rms} & \colhead{$a$} & \colhead{$b$} & \colhead{rms} &\colhead{range for 2nd/3rd}
}
\startdata
\set{O24\_ML4h\_RA} & 0.014 & 0.00 & $3.546$ & $3.028$ & $0.0083$ & $3.720$ & $3.524$ & $0.0015$ & $3.752$ & $3.565$ & $0.0147$ & (0.70,\,0.90) \\
\set{O24\_ML4h\_RA} & 0.006 & 0.00 & $3.483$ & $3.126$ & $0.0078$ & $3.522$ & $3.651$ & $0.0061$ & $3.394$ & $3.681$ & $0.0107$ & (0.60,\,0.89) \\
\set{O24\_ML4h\_RA} & 0.002 & 0.00 & $3.421$ & $3.240$ & $0.0119$ & $3.285$ & $3.739$ & $0.0164$ & $3.284$ & $3.797$ & $0.0148$ & (0.47,\,0.90) \\
\set{O24\_ML4h\_RA} & 0.014 & 0.30 & $3.606$ & $3.031$ & $0.0167$ & $3.689$ & $3.528$ & $0.0035$ & $3.600$ & $3.581$ & $0.0041$ & (0.74,\,0.89) \\
\set{O24\_ML4h\_RA} & 0.006 & 0.30 & $3.525$ & $3.132$ & $0.0114$ & $3.489$ & $3.658$ & $0.0081$ & $3.421$ & $3.688$ & $0.0106$ & (0.60,\,0.89) \\
\set{O24\_ML4h\_RA} & 0.002 & 0.30 & $3.468$ & $3.245$ & $0.0154$ & $3.256$ & $3.745$ & $0.0188$ & $3.286$ & $3.804$ & $0.0158$ & (0.47,\,0.90) \\
\set{O24\_ML4h\_RA} & 0.014 & 0.65 & $3.557$ & $3.048$ & $0.0090$ & $3.630$ & $3.555$ & $0.0084$ & $3.678$ & $3.591$ & $0.0064$ & (0.74,\,0.89) \\
\set{O24\_ML4h\_RA} & 0.006 & 0.65 & $3.540$ & $3.147$ & $0.0138$ & $3.389$ & $3.673$ & $0.0095$ & $3.382$ & $3.705$ & $0.0070$ & (0.65,\,0.89) \\
\set{O24\_ML4h\_RA} & 0.002 & 0.65 & $3.492$ & $3.263$ & $0.0164$ & $3.218$ & $3.758$ & $0.0196$ & $3.353$ & $3.832$ & $0.0125$ & (0.47,\,0.90) \\
\hline
\set{O24\_ML8h\_RA} & 0.014 & 0.65 & $3.655$ & $3.043$ & $0.0257$ & $3.659$ & $3.562$ & $0.0094$ & $3.739$ & $3.598$ & $0.0067$ & (0.73,\,0.89) \\
\set{O24\_ML8h\_RA} & 0.006 & 0.65 & $3.541$ & $3.147$ & $0.0139$ & $3.465$ & $3.681$ & $0.0098$ & $3.447$ & $3.716$ & $0.0063$ & (0.65,\,0.89) \\
\set{O24\_ML8h\_RA} & 0.002 & 0.65 & $3.488$ & $3.264$ & $0.0157$ & $3.221$ & $3.765$ & $0.0212$ & $3.336$ & $3.837$ & $0.0140$ & (0.47,\,0.89) \\
\hline
\set{O14\_ML4h\_RA} & 0.014 & 0.30 & $3.456$ & $2.920$ & $0.0073$ & $3.719$ & $3.404$ & $0.0061$ & $3.576$ & $3.486$ & $0.0085$ & (0.70,\,0.90) \\
\set{O14\_ML4h\_RA} & 0.006 & 0.30 & $3.410$ & $3.016$ & $0.0032$ & $3.569$ & $3.550$ & $0.0089$ & $3.392$ & $3.593$ & $0.0058$ & (0.60,\,0.90) \\
\set{O14\_ML4h\_RA} & 0.002 & 0.30 & $3.387$ & $3.132$ & $0.0036$ & $3.292$ & $3.617$ & $0.0094$ & $3.105$ & $3.690$ & $0.0190$ & (0.48,\,0.90) \\
\enddata
 \tablecomments{Adopted centerings in $\log M/\MS$ are 0.699 for the 1st crossing, and 0.778 for the 2nd, and 3rd crossings.}
\end{deluxetable*}

\subsection{Period-Luminosity relation}\label{ssec:pl}

\begin{figure*}
\includegraphics[width=\linewidth]{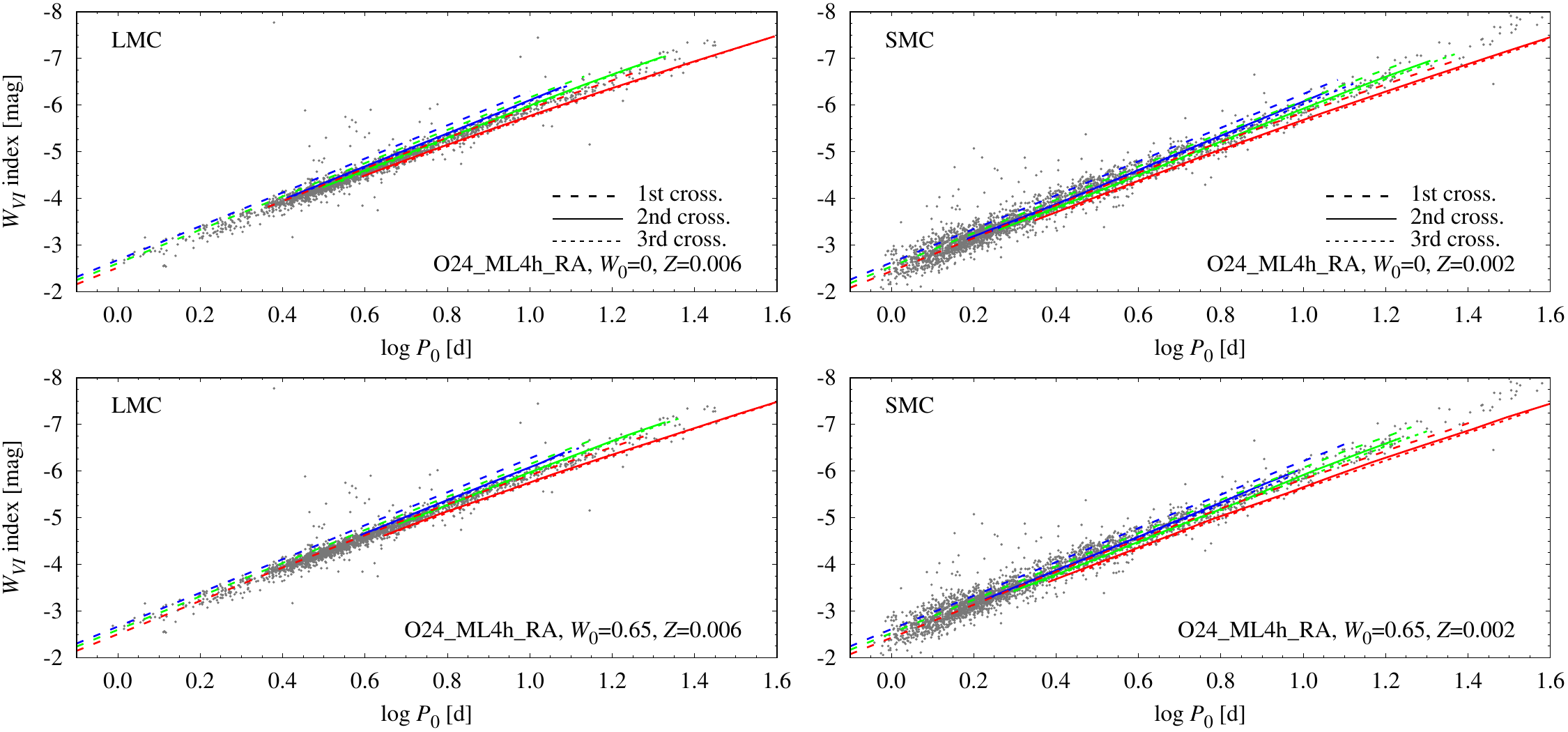}
\caption{Comparison of the observed and model \PL{} relations expressed in terms of the Wesenheit index, $W_{VI}=I-1.55(V-I)$, for the LMC (left) and SMC (right). OGLE data \citep{Soszynski-2015, Soszynski-2017, Soszynski-2019} are compared to $Z=0.006$ (LMC) and $Z=0.002$ (SMC) models of the \set{O24\_ML4h\_RA} set without rotation (top) and including rotation ($W_0=0.65$, bottom). Model relations are shown for the blue and red edges and midline of the IS, with blue, red, and green colors, respectively. Relations for different crossings are plotted with different line styles, as indicated in the legend (relations for the 2nd and 3rd crossings largely overlap). Adopted distance moduli are from \cite{Pietrzynski-2019} for the LMC and from \cite{Graczyk-2020} for the SMC.
\label{fig:pl}}
\end{figure*}

As apparent from Fig.~\ref{fig:hrd_overview} and Tab.~\ref{tab:overview}, independent of the initial rotation rate, inclusion of rotation does not affect the luminosities of the blue loops beyond small, of the order of a few hundredths of a dex shifts. Consequently, the impact of rotation on \PL{} relations should be negligible. In Fig.~\ref{fig:pl}, we compare the \PL{} relations for models of the \set{O24\_ML4h\_RA} set ($\fcor=0.02$, $\fenv=0.04$, $\etar=0.4$), without rotation (top panels) and including rotation (with the largest considered initial rotation rate, $W_0=0.65$; bottom panels), with Magellanic Cloud data from the Optical Gravitational Lensing Experiment, OGLE \citep{Soszynski-2015, Soszynski-2017, Soszynski-2019}. The reddening-free Wesenheit index is used as a luminosity indicator. The only difference that one may spot between the top panels (no rotation) and bottom panels (including rotation) is the different ranges of pulsation periods over which the relations are plotted, resulting from slightly different loop extents in rotating and non-rotating models. Shifts between corresponding relations in the top and bottom panels are smaller than the line widths used in the figure.

On a more quantitative basis, we note that the slopes of the two families of \PL{} relations, with and without rotation, do agree within their errors.

As rotation does not impact the \PL{} relations in a noticeable way, we refrain from further discussion and comparison with observations, and refer the reader to Sect.~3.8 of \citetalias{Smolec-2026} for a detailed discussion of \PL{} relations in non-rotating models, including their metallicity dependence.

For completeness, in Tab.~\ref{tab:pl} we provide coefficients of analytical fits to eq.~\eqref{eq:simplefit} for the \PL{} relation in the $W_{VI}$ Wesenheit index, computed along the midline of the IS for the \set{O24\_ML4h\_RA} set. For $Z=0.006$ and $0.002$, these relations for $W_0=0$ and $W_0=0.65$ correspond to the green lines in Fig.~\ref{fig:pl}. For more detailed comparisons with observations, including different model sets, pass bands, initial rotation rates, or along the edges of the IS, we recommend using the tabular data (see Tab.~\ref{tab:datatab_content}) directly.

\begin{deluxetable*}{lllrrrrrrrrrr}
\tablecaption{Coefficients of the $\PL$ relations as given by eq.~\eqref{eq:simplefit}, computed along the midline of the IS. The first three columns identify the data set, metallicity, and initial rotation rate. Then coefficients and rms follow, separately for the three crossings of the IS. In the last column, we provide the parameter range over which the 2nd and 3rd crossing relations were derived.\label{tab:pl}}
\tablehead{
 & & & \multicolumn{3}{c}{1st cross.} & \multicolumn{3}{c}{2nd cross.} & \multicolumn{3}{c}{3rd cross.} & \\
\colhead{data} & \colhead{$Z$} & \colhead{$W_0$} & \colhead{$a$} & \colhead{$b$} & \colhead{rms} & \colhead{$a$} & \colhead{$b$} & \colhead{rms} & \colhead{$a$} & \colhead{$b$} & \colhead{rms} &\colhead{range for 2nd/3rd}
}
\startdata
\set{O24\_ML4h\_RA} & 0.014 & 0.00 & $-3.552$ & $-4.655$ & $0.006$ & $-3.343$ & $-6.044$ & $0.016$ & $-3.318$ & $-6.018$ & $0.010$ & (0.66,\,1.34) \\
\set{O24\_ML4h\_RA} & 0.006 & 0.00 & $-3.553$ & $-4.588$ & $0.010$ & $-3.405$ & $-5.959$ & $0.017$ & $-3.411$ & $-5.934$ & $0.015$ & (0.50,\,1.34) \\
\set{O24\_ML4h\_RA} & 0.002 & 0.00 & $-3.533$ & $-4.514$ & $0.016$ & $-3.489$ & $-5.895$ & $0.015$ & $-3.448$ & $-5.841$ & $0.017$ & (0.26,\,1.37) \\
\set{O24\_ML4h\_RA} & 0.014 & 0.30 & $-3.557$ & $-4.650$ & $0.008$ & $-3.313$ & $-6.045$ & $0.012$ & $-3.293$ & $-6.018$ & $0.011$ & (0.79,\,1.34) \\
\set{O24\_ML4h\_RA} & 0.006 & 0.30 & $-3.560$ & $-4.582$ & $0.012$ & $-3.409$ & $-5.953$ & $0.017$ & $-3.402$ & $-5.930$ & $0.015$ & (0.51,\,1.34) \\
\set{O24\_ML4h\_RA} & 0.002 & 0.30 & $-3.538$ & $-4.507$ & $0.018$ & $-3.491$ & $-5.888$ & $0.017$ & $-3.444$ & $-5.836$ & $0.017$ & (0.27,\,1.38) \\
\set{O24\_ML4h\_RA} & 0.014 & 0.65 & $-3.547$ & $-4.640$ & $0.007$ & $-3.298$ & $-6.034$ & $0.015$ & $-3.270$ & $-6.011$ & $0.011$ & (0.82,\,1.35) \\
\set{O24\_ML4h\_RA} & 0.006 & 0.65 & $-3.555$ & $-4.571$ & $0.011$ & $-3.387$ & $-5.947$ & $0.014$ & $-3.357$ & $-5.924$ & $0.014$ & (0.68,\,1.36) \\
\set{O24\_ML4h\_RA} & 0.002 & 0.65 & $-3.527$ & $-4.494$ & $0.022$ & $-3.518$ & $-5.884$ & $0.013$ & $-3.454$ & $-5.827$ & $0.014$ & (0.30,\,1.31) \\
\enddata
\tablecomments{Adopted centerings in $\log P$ are 0.556 for the 1st crossing, and 0.997 for the 2nd, and 3rd crossings.}
\end{deluxetable*}

\subsection{Period-Radius relation}\label{ssec:pr}

\begin{figure*}
\includegraphics[width=\linewidth]{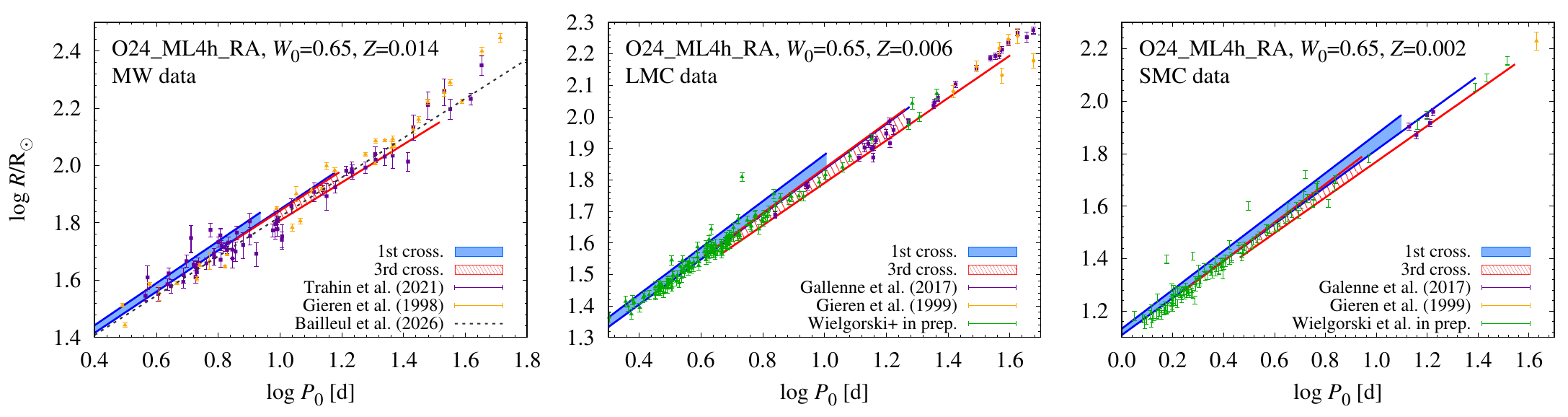}
\caption{$\PR$ relations for \set{O24\_ML4h\_RA} models ($\fcor=0.02$, $\fenv=0.04$, $\etar=0.4$) and the highest initial rotation rate of $W_0=0.65$, for MW, LMC, and SMC metallicities ($Z=0.014$, $0.006$, and $0.002$ in the left, middle, and right panels, respectively), confronted with determinations from \cite{Trahin-2021, Gieren-1998, Gieren-1999, Gallenne-2017} and Wielg\'orski et al. (in prep.). In the first panel, we also display the \PR{} relation from \cite{Bailleul-2026} based on interferometric observations of seven MW Cepheids (dotted gray line).
\label{fig:pr}}
\end{figure*}

Since rotation barely affects the \ML{} and $\PL$ relations, the same is expected for the \PR{} relation. Indeed, \PR{} relations derived for fixed overshooting parameters and different initial rotation rates, including the non-rotating case, nearly overlap. In Fig.~\ref{fig:pr}, we display the \PR{} relations for \set{O24\_ML4h\_RA} models ($\fcor=0.02$, $\fenv=0.04$, $\etar=0.4$) and the largest initial rotation rate in our grids, $W_0=0.65$. As the pulsation period varies considerably across the IS, \PR{} relations are derived separately for the blue and red edges of the IS and, in Fig.~\ref{fig:pr}, are represented with bands, in which the upper envelope corresponds to the blue edge and the lower envelope corresponds to the red edge. Since relations for the 2nd and 3rd crossings nearly overlap, for clarity, we display the latter only. Coefficients of the analytical fit to eq.~\eqref{eq:simplefit} are collected in Tab.~\ref{tab:pr} for the \set{O24\_ML4h\_RA} model set (displayed in the figure) and initial rotation rates of $W_0=0.0,\,0.3,\,0.65$. Coefficients correspond to relations computed along the midline of the IS, which can be considered average relations for a given crossing.

The relations in Fig.~\ref{fig:pr} can be compared with those in the bottom panels of Fig.~14 in \citetalias{Smolec-2026}, where we showed relations for the same overshooting parameters, but without rotation. The only differences can be noticed in terms of the pulsation period range covered by the relations, which is somewhat narrower for models including rotation. We note, however, that relations displayed in \citetalias{Smolec-2026} are based on metallicity-dependent fits that take into account a large grid of 11 metallicities. On a more quantitative basis, we note that the slopes of \PR{} relations for the \set{O24\_ML4h\_RA} model set with $W_0=0$ (no rotation) and $W_0=0.65$ in general match very well for all crossings and metallicities considered.

Since rotation, even with the highest considered initial rotation rate, has a very small impact on the \PR{} relations, comparison with observations yields the same results as in \citetalias{Smolec-2026}. Models including MS core overshooting of $\fcor=0.02$ represent a much better match to observations than models that do not include overshooting. Overshooting cannot be compensated with increasing rotation rate, as the effects of rotation on the blue loop phase are very modest in terms of physical parameters of the models: luminosities and radii are barely different from a non-rotating case. We note that a more detailed comparison with observations for \PR{} relations determined in \citetalias{Smolec-2026}, and those presented here for rotating models, will be conducted in Wielg\'orski et al., in prep.

\begin{deluxetable*}{lllrrrrrrrrrr}
\tablecaption{Coefficients of the $\PR$ relations as given by eq.~\eqref{eq:simplefit}, computed along the midline of the IS. The first three columns identify the data set, metallicity, and initial rotation rate. Then coefficients and rms follow, separately for the three crossings of the IS. In the last column, we provide the parameter range over which the 2nd and 3rd crossing relations were derived.\label{tab:pr}}
\tablehead{
 & & & \multicolumn{3}{c}{1st cross.} & \multicolumn{3}{c}{2nd cross.} & \multicolumn{3}{c}{3rd cross.} & \\
\colhead{data} & \colhead{$Z$} & \colhead{$W_0$} & \colhead{$a$} & \colhead{$b$} & \colhead{rms} & \colhead{$a$} & \colhead{$b$} & \colhead{rms} & \colhead{$a$} & \colhead{$b$} & \colhead{rms} &\colhead{range for 2nd/3rd}
}
\startdata
\set{O24\_ML4h\_RA} & 0.014 & 0.00 & 0.726 & 1.546 & 0.0024 & 0.689 & 1.830 & 0.0015 & 0.685 & 1.825 & 0.0007 & (0.66,\,1.34) \\
\set{O24\_ML4h\_RA} & 0.006 & 0.00 & 0.727 & 1.539 & 0.0030 & 0.697 & 1.820 & 0.0017 & 0.699 & 1.815 & 0.0010 & (0.50,\,1.34) \\
\set{O24\_ML4h\_RA} & 0.002 & 0.00 & 0.723 & 1.529 & 0.0035 & 0.712 & 1.812 & 0.0018 & 0.704 & 1.801 & 0.0021 & (0.26,\,1.37) \\
\set{O24\_ML4h\_RA} & 0.014 & 0.30 & 0.728 & 1.545 & 0.0025 & 0.687 & 1.830 & 0.0009 & 0.686 & 1.824 & 0.0005 & (0.79,\,1.34) \\
\set{O24\_ML4h\_RA} & 0.006 & 0.30 & 0.729 & 1.537 & 0.0036 & 0.698 & 1.819 & 0.0016 & 0.697 & 1.814 & 0.0011 & (0.51,\,1.34) \\
\set{O24\_ML4h\_RA} & 0.002 & 0.30 & 0.726 & 1.527 & 0.0044 & 0.712 & 1.810 & 0.0022 & 0.703 & 1.800 & 0.0021 & (0.27,\,1.38) \\
\set{O24\_ML4h\_RA} & 0.014 & 0.65 & 0.725 & 1.543 & 0.0025 & 0.686 & 1.827 & 0.0017 & 0.682 & 1.822 & 0.0005 & (0.82,\,1.35) \\
\set{O24\_ML4h\_RA} & 0.006 & 0.65 & 0.728 & 1.535 & 0.0035 & 0.697 & 1.817 & 0.0015 & 0.692 & 1.812 & 0.0010 & (0.68,\,1.36) \\
\set{O24\_ML4h\_RA} & 0.002 & 0.65 & 0.723 & 1.525 & 0.0050 & 0.715 & 1.809 & 0.0022 & 0.703 & 1.798 & 0.0020 & (0.30,\,1.31) \\
\enddata
\tablecomments{Adopted centerings in $\log P$ are 0.556 for the 1st crossing, and 0.997 for the 2nd, and 3rd crossings.}
\end{deluxetable*}

\subsection{Period-Age relation}\label{ssec:pa}

\begin{figure*}
\includegraphics[width=\linewidth]{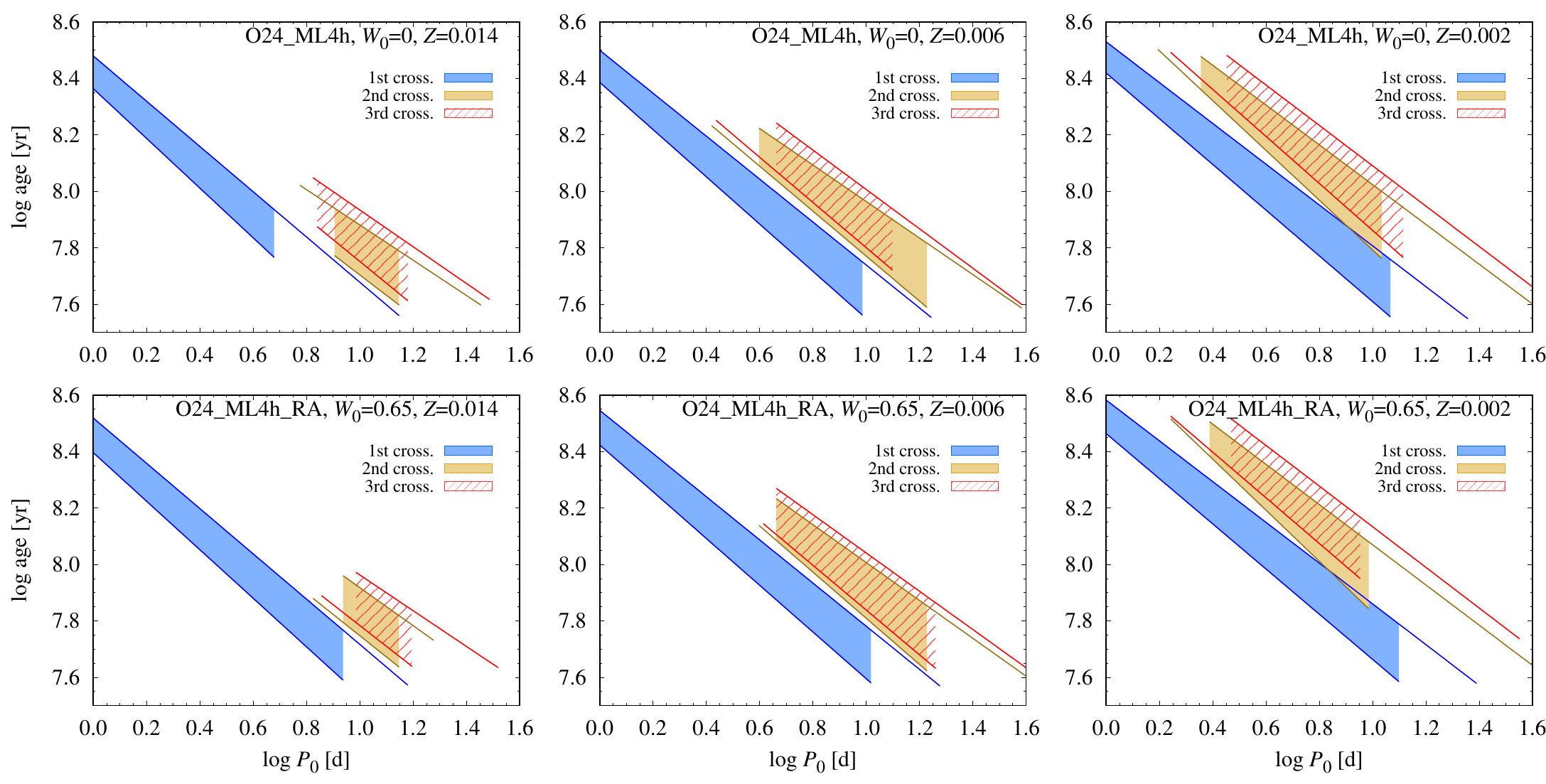}
\caption{$\PA$ relations for \set{O24\_ML4h\_RA} models ($\fcor=0.02$, $\fenv=0.04$, $\etar=0.4$) without rotation (top panels) and with the highest initial rotation rate of $W_0=0.65$ (bottom panels), for MW, LMC, and SMC metallicities ($Z=0.014$, $0.006$, and $0.002$ in the left, middle, and right panels, respectively). Relations are plotted for each crossing; the upper and lower envelopes of the bands correspond to the red and blue edges of the IS, respectively.
\label{fig:pa}}
\end{figure*}

\PA{} relations are expected to be affected by rotation due to increased MS lifetime. This increase is, however, very modest, as we discussed in Sect.~\ref{ssec:blueloop_overview} and quantified in Tab.~\ref{tab:overview} for selected models. Even for the highest considered initial rotation rate, the increase in MS lifetime is only on the order of 10\,\%. This is further visualised with the \PA{} relations, which we display in Fig.~\ref{fig:pa} for a \set{O24\_ML4h\_RA} model set ($\fcor=0.02$, $\fenv=0.04$, $\etar=0.4$). In the top panel, we show relations for non-rotating models as a reference. These largely overlap with the relations we have plotted in Fig.~15 in \citetalias{Smolec-2026}, small differences resulting e.g., from the inclusion of mass loss, also during the MS, in the models presented here. In the bottom panel, we show relations for the largest initial rotation rate, $W_0=0.65$. Just as for the \PR{} relation, relations for the three crossings of the IS are represented with bands, with the upper and lower envelopes corresponding to the red and blue edges of the IS, respectively. Coefficients of the analytical fit to eq.~\eqref{eq:simplefit} are collected in Tab.~\ref{tab:pa} for the \set{O24\_ML4h\_RA} model set (displayed in the figure) and initial rotation rates of $W_0=0.0,\,0.3,\,0.65$. Coefficients correspond to relations computed along the midline of the IS, which can be considered average relations for a given crossing.

Comparing the two cases, $W_0=0$ (top panels) and $W_0=0.65$ (bottom panels), a shift towards longer ages for the latter, at a given pulsation period, for all crossings, is easily noticeable but small. Considering metallicity dependence, both for models without rotation and for models including rotation, the lower the metallicity, the older the Cepheid at a given pulsation period.

We do not include a comparison with other theoretical relations based on non-rotating models, as this was presented in \citetalias{Smolec-2026}. As compared with relations based on Geneva models including rotation, the differences are large and will be discussed in Sect.~\ref{ssec:compgeneva}.

\begin{deluxetable*}{lllrrrrrrrrrr}
\tablecaption{Coefficients of the $\PA$ relations as given by eq.~\eqref{eq:simplefit}, computed along the midline of the IS. The first three columns identify the data set, metallicity, and initial rotation rate. Then coefficients and rms follow, separately for the three crossings of the IS. In the last column, we provide the parameter range over which the 2nd and 3rd crossing relations were derived.\label{tab:pa}}
\tablehead{
 & & & \multicolumn{3}{c}{1st cross.} & \multicolumn{3}{c}{2nd cross.} & \multicolumn{3}{c}{3rd cross.} & \\
\colhead{data} & \colhead{$Z$} & \colhead{$W_0$} & \colhead{$a$} & \colhead{$b$} & \colhead{rms} & \colhead{$a$} & \colhead{$b$} & \colhead{rms} & \colhead{$a$} & \colhead{$b$} & \colhead{rms} &\colhead{range for 2nd/3rd}
}
\startdata
\set{O24\_ML4h\_RA} & 0.014 & 0.00 & $-0.833$ & 7.960 & 0.0125 & $-0.692$ & 7.802 & 0.0060 & $-0.669$ & 7.844 & 0.0087 & (0.66,\,1.34) \\
\set{O24\_ML4h\_RA} & 0.006 & 0.00 & $-0.796$ & 8.001 & 0.0133 & $-0.727$ & 7.873 & 0.0041 & $-0.755$ & 7.910 & 0.0018 & (0.50,\,1.34) \\
\set{O24\_ML4h\_RA} & 0.002 & 0.00 & $-0.761$ & 8.051 & 0.0106 & $-0.796$ & 7.914 & 0.0100 & $-0.771$ & 7.983 & 0.0053 & (0.26,\,1.37) \\
\set{O24\_ML4h\_RA} & 0.014 & 0.30 & $-0.837$ & 7.971 & 0.0120 & $-0.689$ & 7.810 & 0.0030 & $-0.700$ & 7.859 & 0.0008 & (0.79,\,1.34) \\
\set{O24\_ML4h\_RA} & 0.006 & 0.30 & $-0.807$ & 8.015 & 0.0159 & $-0.738$ & 7.885 & 0.0046 & $-0.747$ & 7.920 & 0.0022 & (0.51,\,1.34) \\
\set{O24\_ML4h\_RA} & 0.002 & 0.30 & $-0.772$ & 8.068 & 0.0140 & $-0.801$ & 7.929 & 0.0135 & $-0.768$ & 7.994 & 0.0055 & (0.27,\,1.38) \\
\set{O24\_ML4h\_RA} & 0.014 & 0.65 & $-0.825$ & 7.998 & 0.0124 & $-0.697$ & 7.842 & 0.0090 & $-0.677$ & 7.879 & 0.0023 & (0.82,\,1.35) \\
\set{O24\_ML4h\_RA} & 0.006 & 0.65 & $-0.797$ & 8.044 & 0.0142 & $-0.745$ & 7.913 & 0.0079 & $-0.732$ & 7.946 & 0.0038 & (0.68,\,1.36) \\
\set{O24\_ML4h\_RA} & 0.002 & 0.65 & $-0.758$ & 8.103 & 0.0155 & $-0.814$ & 7.960 & 0.0141 & $-0.761$ & 8.027 & 0.0062 & (0.30,\,1.31) \\
\enddata
 \tablecomments{Adopted centerings in $\log P$ are 0.556 for the 1st crossing, and 0.997 for the 2nd, and 3rd crossings.}
\end{deluxetable*}

\section{Comparison with other studies\label{sec:comparison}}

Below we compare our evolutionary tracks with those from \MESA{} Isochrones \& Stellar Tracks, MIST \citep{Dotter-2016,Choi-2016}, which were also computed with \MESA{} (Sect.~\ref{ssec:compmist}), and with the Geneva tracks and analysis by \cite{Anderson-2014, Anderson-2016} (Sect.~\ref{ssec:compgeneva}). We also test the effects of the custom rotational mixing prescription developed by \cite{Martinelli-2025} (Sect.~\ref{ssec:martinelli}).

We note that for classical Cepheids, evolutionary tracks including rotation were also computed by \cite{Miller-2020}, who used the evolutionary code described by \cite{YoonLanger-2005}. This code uses the diffusion approximation for angular momentum transport \citep{YoonLanger-2004}. Their calculations were limited to solar metallicity. While the authors write that for rapid rotation the blue loop behavior is similar to that found by \cite{Anderson-2014}, they do not provide quantitative measures. A comparison of the tracks they display in Figs.~1 and 2 does not show a significant increase in luminosity for tracks including rotation. Since their tracks are not available, we refrain from a more detailed comparison with that work.

\subsection{Comparison with MIST}\label{ssec:compmist}

The first comparison we conduct is with another set of tracks computed with \MESA{}, the widely used MIST \citep{Choi-2016}. Since these tracks were also computed with \MESA{}, although with one of its earlier versions, 7503, the comparison will be limited. It provides an opportunity to check whether other physical processes that we neglected in our tracks, but are included in MIST, may significantly change the properties of tracks with rotation.

For comparison, we use the MIST version 1.2 EEP tracks for $\feh=0.0$, $-0.25$, and $-0.75$, computed without and including rotation, with $W_0=0.4$ (the only value in the MIST library)\footnote{Tracks were downloaded from \url{https://waps.cfa.harvard.edu/MIST/}}. Detailed characteristics of these models are given in Tab.~1 of \cite{Choi-2016}. Here we just note that the MIST models also adopt the A09 reference solar mixture. Overshooting from the convective core  (both on the MS and during core helium burning) is included with $\fcor=\fhe=0.016$ (about $0.18H_p$ in the step overshooting formalism during the MS). For mass loss, similarly as in our calculations, the `Dutch' wind scheme is used during the early evolutionary phases, followed by the Reimers formula on the RGB ($\eta_R=0.1$) and the Blocker formula on the AGB ($\eta_B=0.2$). For rotation, the adopted parameters are the same as in our model sets \set{*\_RA} (see Tab.~\ref{tab:models}), in particular $\fc=1/30$ and $\fmu=0.05$, and the Spruit-Tayler dynamo is not included. MIST adopts an older version of \MESA{}, prior to the \cite{Paxton-2019} improvement in handling the $f_p$ and $f_T$ factors (see Sect.~\ref{ssec:methods_rotation}). The published tracks, however, adopt $W_0=0.4$, which is below the critical rotation rate, and therefore we do not expect this difference to have a significant impact on the results.

The potentially more impactful differences are the inclusion of additional mixing processes in the MIST tracks, due to semiconvection and thermohaline mixing. Atomic diffusion is also included in the MIST tracks. In our study, these mixing processes are neglected.

\begin{figure*}
\includegraphics[width=\linewidth]{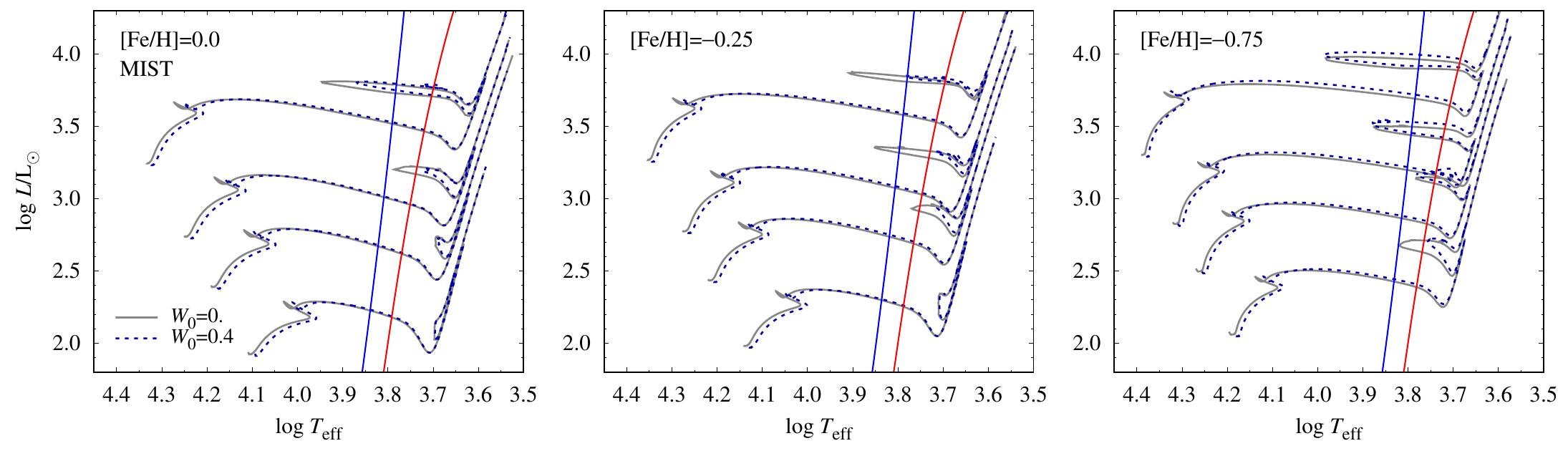}
\caption{Evolutionary tracks for 3, 4, 5, and 7\MS{} from MIST \citep{Choi-2016}. Model metallicities, \feh{}, are $+0.0$, $-0.25$, and $-0.75$ from left to right. Solid lines correspond to models without rotation, while models plotted with blue dashed lines correspond to an initial rotation rate at ZAMS of $W_0=0.4$.
\label{fig:misthr}}
\end{figure*}

In Fig.~\ref{fig:misthr} we plot selected MIST models on HRDs. Quantitative characteristics of specific models, exactly in the form we already presented in Tab.~\ref{tab:overview}, is included in Tab.~\ref{tab:mist}. 

\begin{deluxetable}{rrrrrrrrrrrr}
\tablewidth{0pt}
\tablecaption{The same as Tab.~\ref{tab:overview}, but for selected models from MIST.\label{tab:mist}}
\tablehead{
  &     &  &  &  \multicolumn{2}{c}{1st crossing}  & \multicolumn{2}{c}{2nd crossing}   & \multicolumn{2}{c}{3rd crossing} & \\
\colhead{$M$} & \colhead{[Fe/H]} & \colhead{$W_0$} & \colhead{$\Delta$\,Age$_{\rm MS}$} & \colhead{$\Delta L/\LS$} & \colhead{$v_{\rm rot}$} &  \colhead{$\Delta L/\LS$} & \colhead{$v_{\rm rot}$} &  \colhead{$\Delta L/\LS$} & \colhead{$v_{\rm rot}$} & \colhead{$L_{23}$} & \colhead{$\Delta L_{23,0}$} }
\startdata
3  &  $0.00$ & 0.4 & 2.2 & 0.007 & 88.6 &     - &    - &     - &    -  &     - &  - \\ 
4  &  $0.00$ & 0.4 & 2.5 & 0.008 & 64.4 &     - &    - &     - &    -  &     - &  - \\ 
5  &  $0.00$ & 0.4 & 2.5 & 0.005 & 50.8 &     - &    - &     - &    -  &     - &  - \\ 
7  &  $0.00$ & 0.4 & 3.1 & 0.010 & 32.4 & 0.013 & 52.1 & 0.009 & 49.6  & 0.051 & $-0.004$ \\
3  & $-0.25$ & 0.4 & 2.4 & 0.008 & 80.1 &     - &    - &     - &    -  &     - &  - \\ 
4  & $-0.25$ & 0.4 & 3.5 & 0.014 & 57.3 &     - &    - &     - &    -  &     - &  - \\
5  & $-0.25$ & 0.4 & 3.5 & 0.012 & 44.0 &     - &    - &     - &    -  &     - &  - \\   
7  & $-0.25$ & 0.4 & 3.4 & 0.012 & 29.4 & 0.018 & 55.5 & 0.010 & 48.3  & 0.020 & $-0.008$ \\
3  & $-0.75$ & 0.4 & 4.7 & 0.017 & 68.1 &     - &    - &     - &    -  &     - &  -\\ 
4  & $-0.75$ & 0.4 & 4.6 & 0.017 & 47.6 & 0.020 & 72.8 & 0.025 & 71.3  & 0.023 & $0.005$ \\
5  & $-0.75$ & 0.4 & 5.6 & 0.020 & 36.2 & 0.014 & 57.8 & 0.032 & 53.5  & 0.083 & $0.017$ \\
7  & $-0.75$ & 0.4 & 6.6 & 0.028 & 24.0 & 0.019 & 41.0 & 0.030 & 29.8  & 0.068 & $0.011$ \\ 
\enddata
\end{deluxetable}

We observe no significant differences between the MIST tracks, our results, and their quantitative characteristics. In Fig.~\ref{fig:misthr} we observe qualitatively the same trends in the evolutionary tracks as in Fig.~\ref{fig:hrd_overview}. For rotating models, MS evolution is horizontally shifted towards lower effective temperatures with no significant increase in luminosity. The increase in MS lifetime due to rotation-induced mixing is of the order of a few percent, well matching the increase we recorded in Tab.~\ref{tab:overview} for models of the present study with $W_0=0.3$ and $W_0=0.5$. For the blue loops, rotation-induced increase in luminosity is noticeable but very modest, as quantified in Tab.~\ref{tab:mist}. The extent of the blue loops is similar in MIST models including rotation and those without rotation. Concerning rotational velocities during the crossings of the IS (compare Tabs~\ref{tab:overview} and \ref{tab:mist}), those from MIST are very close to those from the present study and models with $W_0=0.5$. The luminosity extent of the loops, characterized by $L_{23}$ and $\Delta L_{23,0}$, is similar to that found in our models.

The two major issues we have identified with rotation implementation in \MESA{}, very modest rotation induced increase in blue loop luminosity and too large surface rotational velocities during the IS crossings, are also present in MIST models.

\subsection{Comparison with Geneva tracks}\label{ssec:compgeneva}

Geneva evolutionary tracks for the mass and metallicity range considered in this study were published by \cite{Georgy-2013} \citep[see also][for further details]{Ekstrom-2012} and were analyzed in the context of classical Cepheids by \cite{Anderson-2014, Anderson-2016}. As discussed in Sect.~\ref{ssec:methods_rotation}, rotation in these models is implemented using an advective–diffusive scheme. The advective term in the equation for angular momentum transport is included only during MS evolution; afterwards, the scheme is fully diffusive \citep{Nandal-2024}. Still, this is not equivalent to the scheme used in \MESA, as the same diffusion coefficient is adopted for both angular momentum transport and chemical element mixing. In particular, there is no equivalent of the $f_c$ parameter in \MESA, which scales the efficiency of chemical mixing relative to angular momentum transport \citep[see][for implementation details]{Ekstrom-2012}. Geneva models adopt \cite{Asplund-2005} reference solar mixture, which is slightly different from A09 that we adopt -- see \cite{Asplund-2009} for comparison. The models include MS convective core overshooting (step formalism) of $0.1H_p$, both during the MS and core helium burning. For the purpose of this comparison, we have downloaded the tracks from the SYCLIST website\footnote{\url{https://www.unige.ch/sciences/astro/evolution/en/database/syclist}} and selected evolutionary tracks directly included in the database (so they are not interpolated from other tracks). These are 3, 4, 5, and 7\MS{} tracks. The metallicities are the same as in our models, i.e., $Z=0.014$, $0.006$, and $0.002$, we note however, that the adopted reference solar mixture, as well as helium enrichment ($\Delta Y/\Delta Z=1.5$ in our models vs.\ $\Delta Y/\Delta Z=1.257$ in Geneva models) are different. So while these metallicities may be considered representative for MW, LMC, and SMC Cepheids in both studies, the abundances are not one-to-one identical. For post-MS evolution, Geneva models include mass loss following the Reimers formula with \etar=0.5.
 
\cite{Anderson-2016} considered evolutionary tracks with two initial rotation rates, $\omega_0=0.5$ and $\omega_0=0.9$, with the former referred to as an average rotation rate. We remind the reader of the difference in the definitions of the critical rotation rates in \MESA{} and the Geneva codes, discussed in Sect.~\ref{ssec:oocrit}. These two initial rotation rates correspond to our $W_0\approx0.3$ and $W_0\approx0.65$. Most of our comparison will focus on the average rotation, i.e., $\omega_0=0.5$ (Geneva) and $W_0=0.3$ (\MESA{}). Since the extent of the MS core overshooting adopted in the Geneva tracks is $0.1H_p$, for comparison we will use models of \set{O14\_ML4h\_RA}, i.e., with \fcor=0.01, which corresponds to $\approx0.115H_p$ (see relation between step and exponential overshooting parameters derived in the Appendix in \citetalias{Smolec-2026}). We note that this is the model set that was discussed in detail in Sect.~\ref{ssec:blueloop_overview}, for which selected evolutionary tracks are plotted in Fig.~\ref{fig:hrd_overview}, and their quantitative characteristics are collected in Tab.~\ref{tab:overview}.

\begin{figure*}
\includegraphics[width=\linewidth]{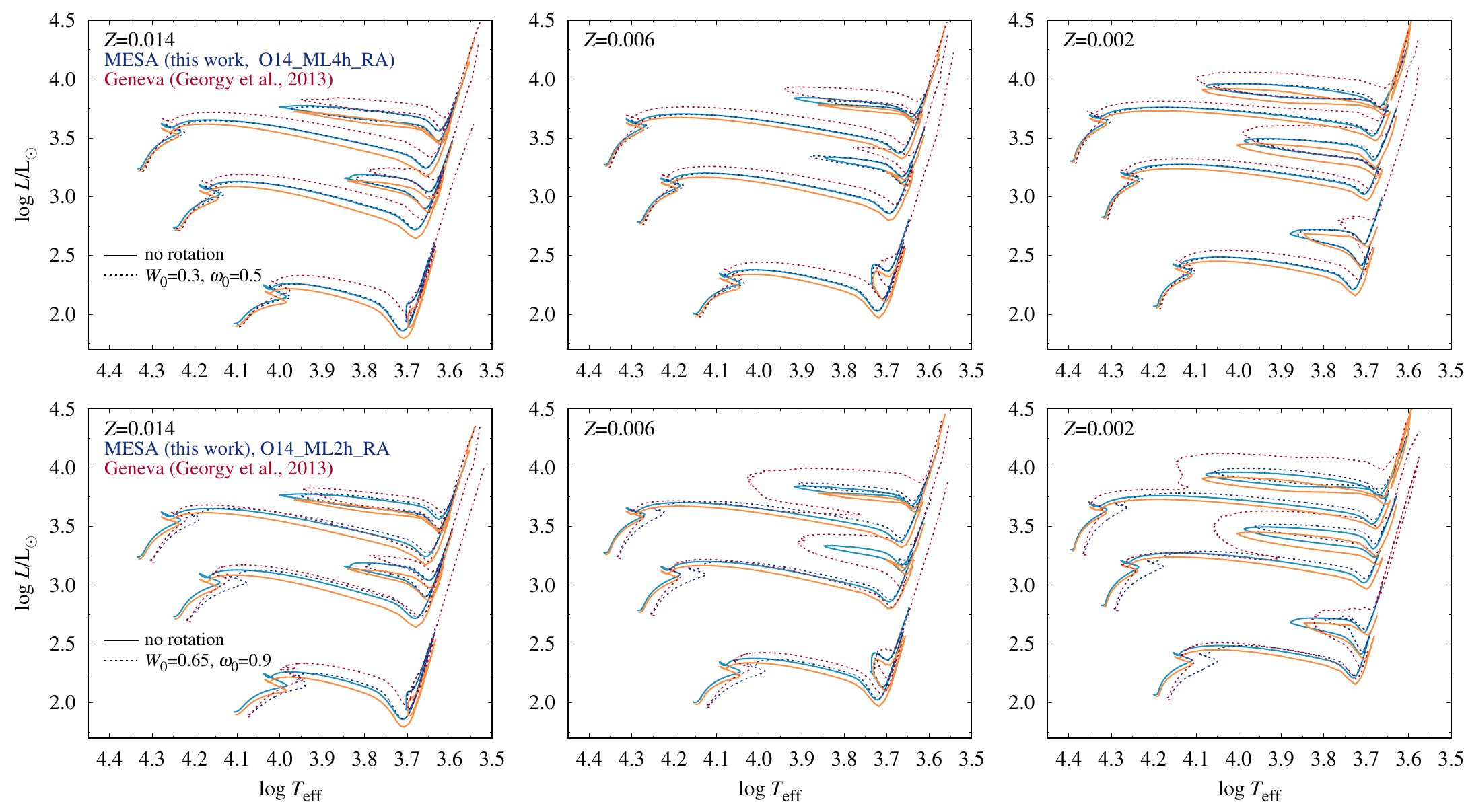}
\caption{Comparison of \MESA{} tracks computed in this study (\set{O14\_ML4h\_RA}; blue lines) with models computed with the Geneva code \citep{Georgy-2013} from the SYCLIST database (orange lines). The 3, 5, and 7\MS{} models are presented with $Z=0.014$ (left), $Z=0.006$ (middle), and $Z=0.002$ (right). Non-rotating models are plotted with solid lines and light color. Models including rotation are plotted with dotted lines and dark color. The initial rotation rates are $W_0=0.3$ and $\omega_0=0.5$ in the top panels, and $W_0=0.65$ and $\omega_0=0.9$ in the bottom panels.
\label{fig:genevahr}}
\end{figure*}

Geneva evolutionary tracks are presented in Fig.~\ref{fig:genevahr}. In the top panels, we show evolutionary tracks for $\omega_0=0.5$, while in the bottom panels we show tracks with $\omega_0=0.9$. For comparison, we over-plot \MESA{} models of the \set{O14\_ML4h\_RA} set with $W_0=0.3$ (top) and $W_0=0.65$ (bottom). For reference, non-rotating tracks are also plotted for both the Geneva code and \MESA{}. For clarity, we show the tracks for 3, 5, and 7\MS{} only. In Tab.~\ref{tab:geneva} we collect the quantitative characteristics of the Geneva tracks, similar to those for the \MESA{} tracks in Tab.~\ref{tab:overview}.

The non-rotating tracks (solid lines in Fig.~\ref{fig:genevahr}) are qualitatively similar. The slight differences in luminosity levels (\MESA{} tracks being slightly brighter) may be attributed to a bit higher MS core overshooting in \MESA{} tracks, and to other differences in the microphysical setup of the models. Comparing rotating tracks (dotted lines in Fig.~\ref{fig:genevahr}), differences are striking and start to be apparent early on, during the MS evolution. For \MESA{} tracks, the MS is horizontally shifted towards lower effective temperatures, without any significant shift in luminosity. Geneva tracks including rotation also start at lower effective temperatures and the shift in \Teff{} is of the same order. During the MS evolution, in contrast to \MESA{} tracks, luminosity increases, as compared with the non-rotating track. The MS hook and the TAMS are located at considerably higher luminosities as compared with tracks without rotation. Considering the MS lifetime (4th column in Tab.~\ref{tab:geneva}), its increase is significantly larger than for \MESA{} tracks and amounts to $\approx 20$\,\% for $\omega_0=0.5$ and to $\approx 25$\,\% for $\omega_0=0.9$, on average.

The following evolutionary stages also proceed at overall higher luminosity levels compared with non-rotating tracks, and with \MESA{} tracks in general. For the blue loops, the scenario depends on mass, metallicity, and initial rotation rate. Quantitative estimates of the rotation-induced increase in luminosity at different crossings of the IS are collected in Tab.~\ref{tab:overview} (\MESA) and in Tab.~\ref{tab:geneva} (Geneva code). For the lower initial rotation rate (top panels in Fig.~\ref{fig:genevahr}), and for $Z=0.014$, and $0.006$, the whole blue loop is brighter in the Geneva tracks as compared with \MESA; the 2nd crossing of the Geneva tracks roughly overlapping with the 3rd crossing of \MESA{} tracks. For $Z=0.002$ (and 5, 7\MS), we observe that the 2nd crossing for \MESA{} and Geneva tracks nearly overlaps. But then, the thickness of the blue loop is significantly larger in the Geneva tracks, leading to the 3rd crossing at a significantly larger luminosity level. The increased thickness of the Geneva blue loops is visible for all tracks in Fig.~\ref{fig:genevahr}, but most significantly for $\omega_0=0.9$ (bottom panels in Fig.~\ref{fig:genevahr}), where for lower metallicities and higher masses the thickness of the loops easily exceeds 0.2\,dex, while in the \MESA{} tracks it stays below 0.05\,dex (see also $L_{23}$ in Tabs~\ref{tab:overview} and \ref{tab:geneva}). For $Z=0.002$, and 5, and 7\MS{} Geneva tracks, the core helium burning appears to start before the post-MS evolution reaches the blue edge of the IS. Consequently, a single crossing of the IS, that physically corresponds to the 3rd crossing, occurs at the highest luminosity levels, as the star evolves directly towards AGB.

The differences in loop thickness are also well captured with the $L_{23}$ and $\Delta L_{23,0}$ parameters in the last columns of Tabs~\ref{tab:overview} and \ref{tab:geneva}. For rotating tracks, unless the loop's turning point is close to the midline (e.g., for 5\MS, $Z=0.014$ models) the loops are strongly extended in luminosity, with $L_{23}$ exceeding 0.2\,dex. The increase with respect to non-rotating tracks is also large, typically exceeding 0.1\,dex, in contrast to tracks computed with \MESA, for which the loop broadening is very modest.

Considering the surface rotational velocity during IS crossings (compare the relevant columns in Tabs~\ref{tab:overview} and \ref{tab:geneva}), Geneva tracks offer a much better agreement with the slow rotation rates observed in classical Cepheids. For their average rotation rate, i.e., $\omega_0=0.5$, velocities are significantly lower than those predicted with \MESA{} tracks. For the 3rd crossing, all values reported in Tab.~\ref{tab:geneva} stay below 25\,km/s. For the 2nd crossing, velocities are somewhat larger and may exceed 25\,km/s. Interestingly, rotation velocities seem to slightly increase with mass in the Geneva tracks, while they decrease with increasing mass for \MESA{} tracks (see Fig.~\ref{fig:vrot_overview}).

Based on this comparison, we may conclude that rotation-induced mixing during the MS evolution must be significantly more efficient in the Geneva tracks, as compared with \MESA{} tracks, leading to a larger convective core, a significant increase in the MS lifetime, to an increase in the luminosity of the blue loops, and to an increase in the thickness of the blue loops. This is further quantified with \ML{} and \PA{} relations.

\begin{figure*}
\includegraphics[width=\linewidth]{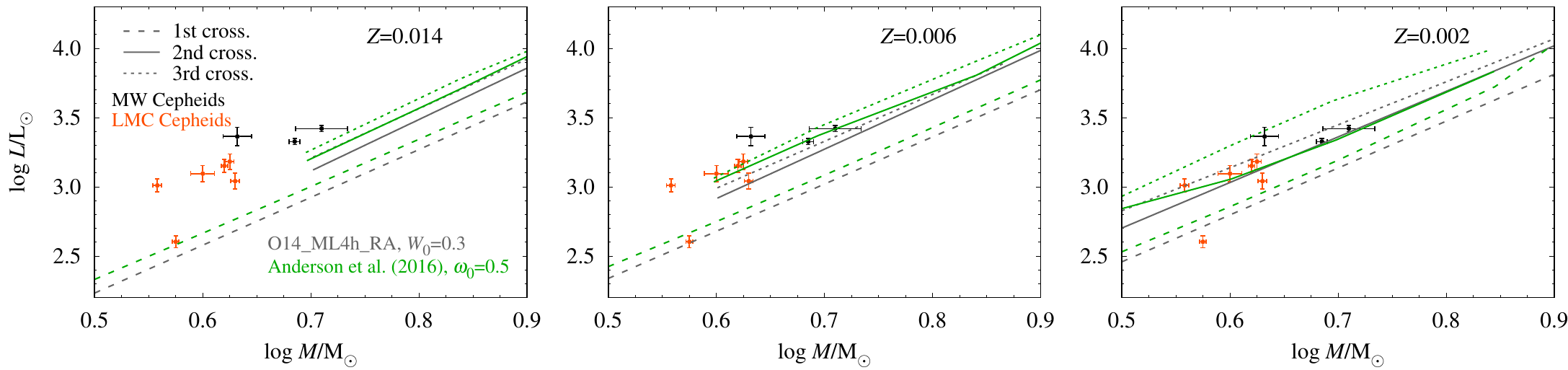}
\caption{$\ML$ relations for \set{O14\_ML4h\_RA} models (\fcor=0.01, \fenv=0.04, \etar=0.4) with $W_0=0.3$ (gray lines) compared with \ML{} relations from \cite{Anderson-2016} ($\beta_{\rm H}=0.1H_p$, $\omega_0=0.5$; green lines). Comparison is presented for $Z=0.014$ (left), $0.006$ (middle), and $0.002$ (right). The theoretical relations are confronted with determinations from \cite{Gallenne-2018-V1334Cyg, Gallenne-2025-SuCyg, Pilecki-2018, Evans-2024-Polaris}.
\label{fig:ml_geneva}}
\end{figure*}

For the following of this comparison, we focus on tracks with $\omega_0=0.5$ and the corresponding $W_0=0.3$ in \MESA. In Fig.~\ref{fig:ml_geneva} we display \ML{} relations. Geneva relations were constructed directly using data from Tabs~A.1--A.3 in \cite{Anderson-2016}. The \MESA{} tracks we use in comparison are from the \set{O14\_ML4h\_RA} set representing the closest match to the Geneva tracks in terms of the adopted MS core overshooting. Coefficients of these relations expressed in the form of eq.~\eqref{eq:simplefit} are collected in Tab.~\ref{tab:ml}. Although both \MESA{} and Geneva tracks were initialized with closely matching initial rotation rates, and adopt a similar extent of MS core overshooting, the corresponding \ML{} relations are significantly different. Relations based on the Geneva tracks predict significantly larger luminosities at a given mass. Since for tracks without rotation, the luminosity levels of the blue loops are similar for both \MESA{} and Geneva models (see also the comparison in the top panels of fig.\ 13 in \citetalias{Smolec-2026}), the difference must be attributed to the different impact of rotation on evolution in the two evolutionary codes.

In Fig.~\ref{fig:ml_geneva}, we observe a nearly constant shift in luminosity for the 1st crossing, Geneva relations are brighter by about 0.08\,dex, independent of metallicity. For the 2nd crossing, the scenario depends on metallicity. For $Z=0.014$, just as for the 1st crossing, we observe a nearly constant (i.e., independent of mass) shift in luminosity, Geneva relations predicting higher luminosities by about 0.08\,dex. Similar for $Z=0.006$, however here the shift is 0.12\,dex, and then decreases to about 0.04\,dex for larger masses. For $Z=0.002$, only for lower masses do the two relations differ, the one based on Geneva tracks predicting larger luminosities. For masses larger than 4\MS{}, the two relations predict the same luminosity. For the 3rd crossing, the relation based on Geneva tracks is always significantly brighter than the relation based on \MESA{} tracks. The difference increases with decreasing metallicity. For $Z=0.014$ the offset is of the order of 0.07\,dex, while for $Z=0.002$ it increases to 0.19\,dex at $\log M/\MS=0.7$. For $Z=0.002$ we also observe the largest separation between the 2nd and 3rd crossings for Geneva tracks, see Fig.~\ref{fig:genevahr} and $L_{23}$ in Tab.~\ref{tab:geneva}. This perfectly matches the scenario outlined in the preceding paragraphs based on the evolutionary tracks themselves plotted in Fig.~\ref{fig:genevahr}. By comparing with Fig.~\ref{fig:ml}, we may conclude that to reach similar luminosity levels with \MESA{} tracks, one needs to increase the MS core overshooting to at least \fcor=0.02, i.e., well above $0.2H_p$ (in the step overshooting formalism), the initial rotation rate having negligible effect on the \ML{} relation.

It is interesting to note that both \MESA{} and Geneva tracks predict a similar minimum mass to enter the IS for all metallicities, as \ML{} relations start at the same mass. We note that this minimum mass may be too high to explain short-period metal-rich Cepheids. As we have demonstrated in \citetalias{Smolec-2026} however, the appearance and extent of the blue loops strongly depends on metallicity and overshooting parameters, and dense model grids, in particular in $Z$, are necessary for more general comparisons and definite conclusions.

\begin{figure*}
\includegraphics[width=\linewidth]{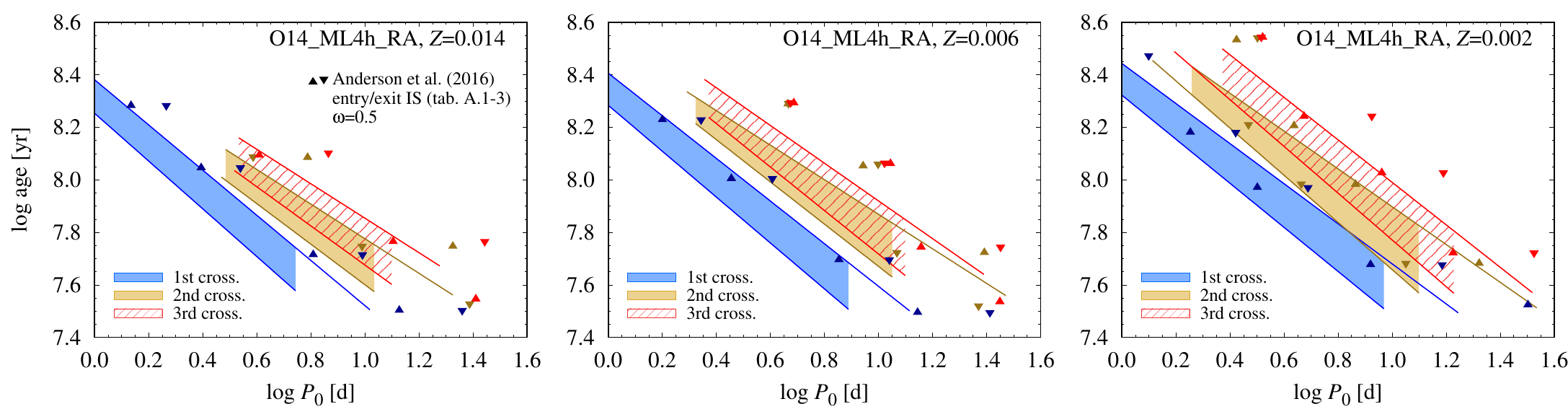}
\caption{$\PA$ relations for \set{O14\_ML4h\_RA} models (\fcor=0.01, \fenv=0.4, \etar=0.4) with $W_0=0.3$ compared with \ML{} relations from \cite{Anderson-2016} ($\beta_{\rm H}=0.1H_p$, $\omega_0=0.5$; triangles). Comparison is presented for $Z=0.014$ (left), $0.006$ (middle), and $0.002$ (right).
\label{fig:pa_geneva}}
\end{figure*}

In Fig.~\ref{fig:pa_geneva}, we compare \PA{} relations for the same set of models as in the comparison of \ML{} relations. For Geneva tracks, again we use data from Tabs~A.1--A.3 of \cite{Anderson-2016}. We observe that for all crossings, Geneva tracks predict significantly older Cepheids at a given pulsation period. This is in agreement with the observation made above (compare also $\Delta{\rm Age}_{\rm MS}$ in Tabs~\ref{tab:overview} and \ref{tab:geneva}), that inclusion of rotation leads to a significantly longer MS lifetime in Geneva tracks. During the 1st crossing, ages are 11--21\,\% longer as compared with \MESA{} tracks initialized with the same surface rotational velocity. For the blue loop (2nd and 3rd crossing), the ages are 8--16\,\% longer. The exact numbers depend on mass and metallicity. Consequently, the determination of Cepheid ages from the \PA{} relation depends not only on whether rotation is included in the model calculations, but also on which code was used to compute the underlying rotating models.

Independent methods for determining Cepheid ages would be valuable for verifying the \PA{} relations. Ages inferred from modeling the co-evolution of Cepheids in binary systems \citep[e.g.][]{Neilson-2012-cep227,Deka-2025} can provide useful constraints, but they are not fully independent, as they also rely on stellar evolutionary calculations and associated model assumptions. Moreover, such systems are scarce. Ages may also be inferred for classical Cepheids in open clusters through isochrone fitting. Here again, the results are model dependent, the sample of Cepheids in open clusters is limited, and the method itself is also more challenging to apply; see e.g., \cite{Medina-2021}.

\begin{deluxetable}{rrrrrrrrrrrr}
\tablewidth{0pt}
\tablecaption{The same as Tab.~\ref{tab:overview}, but for selected models computed with the Geneva code from SYCLIST database \citep{Georgy-2013}.\label{tab:geneva}}
\tablehead{
  &     &  &  &  \multicolumn{2}{c}{1st crossing}  & \multicolumn{2}{c}{2nd crossing}   & \multicolumn{2}{c}{3rd crossing} & \\
\colhead{$M$} & \colhead{$Z$} & \colhead{$\omega_0$} & \colhead{$\Delta$\,Age$_{\rm MS}$} & \colhead{$\Delta L/\LS$} & \colhead{$v_{\rm rot}$} &  \colhead{$\Delta L/\LS$} & \colhead{$v_{\rm rot}$} &  \colhead{$\Delta L/\LS$} & \colhead{$v_{\rm rot}$} & \colhead{$L_{23}$} & \colhead{$\Delta L_{23,0}$}}
\startdata
3  & 0.014 & 0.5 & 25.1 & 0.154 & 46.7 &        - &    - &     - &    -  &     - &     - \\ 
3  & 0.014 & 0.9 & 31.2 & 0.156 & 92.7 &        - &    - &     - &    -  &     - &     - \\ 
3  & 0.006 & 0.5 & 21.4 & 0.141 & 38.9 &        - &    - &     - &    -  &     - &     - \\ 
3  & 0.006 & 0.9 & 25.7 & 0.111 & 78.8 &        - &    - &     - &    -  &     - &     - \\ 
3  & 0.002 & 0.5 & 20.0 & 0.125 & 30.1 &        - &    - &     - &    -  &     - &     - \\ 
3  & 0.002 & 0.9 & 22.7 & 0.063 & 61.0 & $-0.002$ & 22.2 & 0.114 & 16.0  & 0.200 & $0.116$ \\
4  & 0.014 & 0.5 & 23.3 & 0.158 & 32.2 &        - &    - &     - &    -  &     - &     - \\ 
4  & 0.014 & 0.9 & 28.6 & 0.132 & 64.7 &        - &    - &     - &    -  &     - &     - \\ 
4  & 0.006 & 0.5 & 20.2 & 0.136 & 26.9 &        - &    - &     - &    -  &     - &     - \\ 
4  & 0.006 & 0.9 & 23.5 & 0.071 & 55.1 &        - &    - &     - &    -  &     - &     - \\ 
4  & 0.002 & 0.5 & 19.2 & 0.114 & 21.0 &    0.088 & 13.6 & 0.191 &  4.6  & 0.226 & $0.103$ \\
4  & 0.002 & 0.9 & 21.6 & 0.021 & 41.5 & $-0.103$ & 59.7 & 0.168 & 31.1  & 0.394 & $0.271$ \\
5  & 0.014 & 0.5 & 23.7 & 0.163 & 24.4 &    0.118 & 24.6 & 0.113 & 15.3  & 0.039 & $-0.004$ \\
5  & 0.014 & 0.9 & 28.0 & 0.104 & 48.2 &  $0.145$ & 36.9 & 0.124 & 27.1  & 0.022 & $-0.021$ \\
5  & 0.006 & 0.5 & 19.4 & 0.132 & 19.8 &        - & 15.0 &     - & 12.7  & 0.020 &     - \\ 
5  & 0.006 & 0.9 & 21.9 & 0.034 & 40.9 &        - & 44.0 &     - & 16.9  & 0.253 &     - \\ 
5  & 0.002 & 0.5 & 18.3 & 0.101 & 15.3 &    0.057 & 28.5 & 0.201 & 16.9  & 0.255 & $0.143$ \\
5  & 0.002 & 0.9 & 21.7 &     - &    - &        - &    - & 0.195 & 19.9  &     - &     - \\ 
7  & 0.014 & 0.5 & 22.6 & 0.158 & 15.3 &    0.120 & 37.4 & 0.168 & 23.8  & 0.083 & $0.048$ \\
7  & 0.014 & 0.9 & 25.8 & 0.056 & 31.1 &  $0.051$ & 69.3 & 0.130 & 36.0  & 0.114 & $0.079$ \\
7  & 0.006 & 0.5 & 18.3 & 0.127 & 12.2 &    0.083 & 33.5 & 0.148 & 18.0  & 0.094 & $0.064$ \\
7  & 0.006 & 0.9 & 21.0 &     - &    - &        - &    - & 0.214 &  3.0* &     - &     - \\ 
7  & 0.002 & 0.5 & 17.2 & 0.150 &  8.8 &    0.049 & 23.1 & 0.181 & 15.2  & 0.164 & $0.132$ \\
7  & 0.002 & 0.9 & 26.3 &     - &    - &        - &    - & 0.247 & 10.9  &     - &     - \\ 
\enddata
\end{deluxetable}

\subsection{\MESA{} tracks with mixing prescription from \cite{Martinelli-2025}}\label{ssec:martinelli}

\cite{Martinelli-2025} replaced the default \MESA{} diffusion coefficients with a custom rotational mixing prescription during the MS, calibrated to reproduce the transport of elements and the evolution of surface equatorial velocity in the Geneva models of \cite{Georgy-2013}. The calibration was performed for models up to $4\MS{}$, and only for this mass range are the relevant coefficients of the new prescription, which also depend on the initial rotation rate, provided. After MS, the \MESA's default mixing prescription is used. Such modified models yield significantly more efficient mixing during MS evolution. \cite{Martinelli-2025} focused on modeling the color–magnitude diagrams of stellar clusters to study the impact of rotation on extended main-sequence turn-offs and extended red clumps. They did not investigate higher masses or the development of blue loops. 

Since the modified mixing routines and the calibrated parameters of the custom mixing prescription of \cite{Martinelli-2025} are available on Zenodo \citep{martinelli-zenodo}, we attempt to study the impact of this prescription on the blue loops. The parameters of the mixing prescription appear to be sensitive to other model assumptions. An attempt to use the same microphysical and numerical setup as adopted in our models failed, leading to odd evolutionary tracks. Consequently, we decided to retain the original setup of the \cite{Martinelli-2025} models, summarized in their tab.~1\footnote{We also used the same \MESA{} version for this exercise, i.e., 24.03.1.}. A mass of $4\MS$ represents the approximate threshold above which blue loops begin to develop. Since the calibrated parameters are available only up to this mass, we consider a $4\MS$ model with $Z=0.002$, as blue loops develop more readily at low metallicity. Otherwise, the models adopt MS core overshooting in the step formulation with an extent of $0.1H_{\rm p}$. For this exercise we set the initial rotation rate to $W_0=0.3$

\begin{figure}
\includegraphics[width=\linewidth]{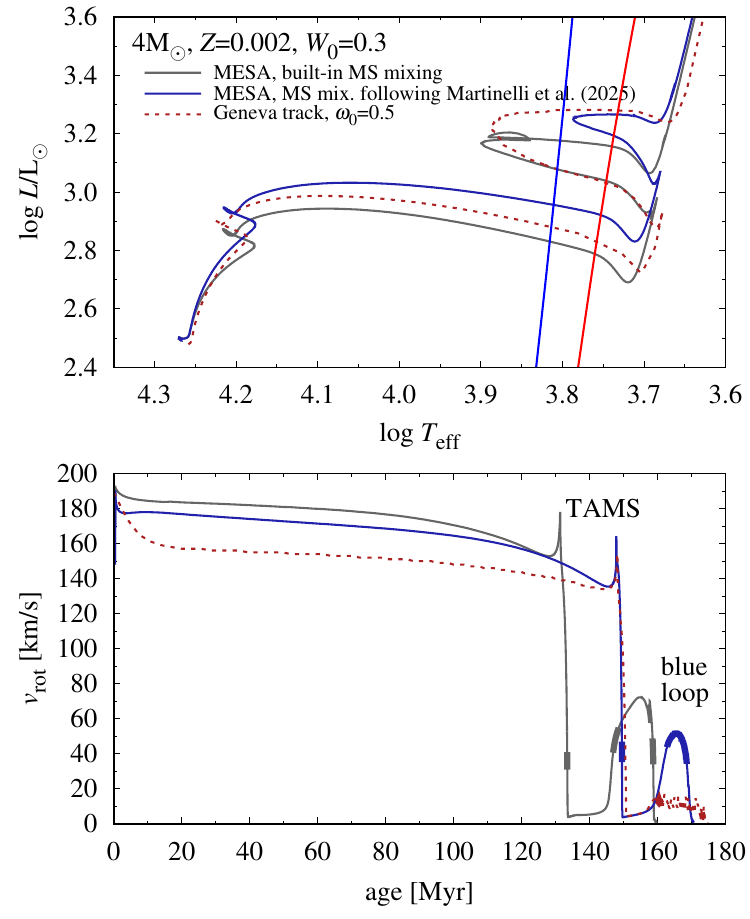}
\caption{HRD (top) and evolution of surface rotational velocity (bottom) for 4\MS, $Z=0.002$, $W_0=0.3$ model computed with \MESA's default rotational mixing prescription (solid gray line) and modified prescription from \cite{Martinelli-2025}, calibrated to emulate the rotational induced mixing in the Geneva code during MS. Corresponding Geneva track ($\omega_0=0.5$) is plotted with dashed red line for comparison. In the bottom panel, thick line segments correspond to three crossings of the IS.
\label{fig:martinelli}}
\end{figure}

In the top panel of Fig.~\ref{fig:martinelli} we display evolutionary tracks computed with the \MESA's default rotational mixing prescription (solid gray line) and the modified prescription from \cite{Martinelli-2025} (solid blue line). In addition, we over-plot Geneva evolutionary track of 4\MS{}, $Z=0.002$, and $\omega_0=0.5$. In the bottom panel we show the evolution of surface rotational velocity in the three models. Thick line sections correspond to crossings of the IS.

We observe that with custom mixing prescription, starting from ZAMS, evolution proceeds at higher luminosities compared with the \MESA's default rotational mixing. The entire blue loop is significantly brighter, with 2nd crossing placed above 3rd crossing of the model with default rotational mixing. In the bottom panel of Fig.~\ref{fig:martinelli} we may observe that the MS lifetime in model adopting custom rotational prescription matches the MS lifetime of the Geneva track. 

Surface rotational velocities during the IS crossings (thick line segments in the bottom panel of Fig.~\ref{fig:martinelli}) are little affected by the custom prescription, i.e., they roughly match the ranges in model adopting \MESA's default rotational mixing.
In the Geneva track, surface rotation velocities drop much more efficiently, already at the beginning of the MS. Since the custom prescription still uses the fully diffusive scheme for angular momentum transport, it appears that this stronger decrease of the surface rotational velocities in the Geneva code is intrinsically linked to the different, advective-diffusive scheme. After the MS, the Geneva code also switches to a fully diffusive scheme \citep{Nandal-2024}, but the efficiency of element mixing corresponds to that of angular momentum transport and is hence significantly higher than in \MESA{} tracks, which may be another factor contributing to lower rotational velocities during the blue loop phase.

Interestingly, the evolutionary track based on mixing prescription calibrated to emulate the Geneva tracks is qualitatively different from the Geneva track of the corresponding physical parameters. While luminosity at the 3rd crossing matches in both models, the blue loop is significantly more extended and more thick in the original Geneva track. Also, the evolution of surface rotational velocity proceeds differently, with lower rotational velocities during the MS evolution in Geneva track, and significantly lower velocities during core helium burning. One has to keep in mind, that we are at the edge of the calibration conducted by \cite{Martinelli-2025}, at the highest considered mass. The calibration was also conducted using evolutionary tracks of solar metallicity ($Z=0.014$), while our model is metal poor in order to develop the loop. Transport of angular momentum is still treated in a fully diffusive manner during the MS evolution.

This exercise confirms that the modest effects of rotation on the appearance of the blue loops in \MESA{} tracks results from inefficient (compared to the Geneva code) MS mixing.

\section{Summary and conclusions}\label{sec:summary}

\cite{Anderson-2014, Anderson-2016} studied rotating Cepheid models using evolutionary tracks computed with the Geneva code \citep{Georgy-2013}, yielding very interesting results. Rotation-induced mixing during the MS evolution elevates the luminosity level during subsequent evolution, in particular during core helium burning, when the star crosses the IS and becomes a Cepheid. Individual loops expand significantly in the vertical direction, leading to an increased separation between the 2nd and 3rd crossings. These two factors contribute to alleviating the mass discrepancy problem, without the need to significantly increase MS core overshooting. By significantly increasing the MS lifetime, the \PA{} relation is altered, leading to older Cepheids at a given pulsation period (by about 20\,\%) and thus impacting inferences in galactic archaeology studies \citep[see e.g.,][]{Skowron-2019}. Rotation velocities during core helium burning are also in reasonable agreement with observations.

There is no unique way to implement rotation in stellar models. The two main approaches differ in the way angular momentum transport due to meridional circulation is treated. In the first approach, adopted in the Geneva code, it is implemented as an advective process. In the second, adopted e.g., in \MESA{}, it is treated using a diffusive scheme (see Sect.~\ref{ssec:methods_rotation}). For the remaining rotation-induced processes, diffusive equations are used in both methods.

While the advective–diffusive scheme offers a physically more consistent picture of rotation-induced mixing \citep[see e.g.,][]{Maeder-2009book}, the fully diffusive approach is widely used. In our work we have investigated whether it leads to results similar to those obtained with the advective–diffusive scheme. To this aim we computed a grid of models including rotation with \MESA{}. The most important findings are:
\begin{itemize}
\item[--] Independent of the initial rotation rate, the increase in luminosity of the blue loops is very modest ($<0.04$\,dex in $\log L/\LS$).

\item[--] The vertical extent of the loops is barely affected compared with non-rotating models.

\item[--] As a consequence, inclusion of rotation alone, without a significant increase in MS core overshooting, does not offer a solution to the mass discrepancy problem.

\item[--] The period–radius and period–luminosity relations barely differ from their non-rotating counterparts. While we provide \PR{} and \PL{} relations for rotating models, they are very similar to those based on non-rotating models, which we studied in detail in \citetalias{Smolec-2026}.

\item[--] The increase in MS lifetime is also very modest, with the resulting \PA{} relation implying Cepheids only a few per cent older at a given pulsation period compared with the relation based on non-rotating tracks.

\item[--] The predicted surface rotational velocities at all crossings of the IS are too large ($>20$\,km/s) and cannot be reconciled with observations, except for the most massive, 8\MS{} models.

\end{itemize}
Rotation has several free parameters, which we explored to some degree (Sect.~\ref{ssec:freep}), but the results outlined above are robust and not affected by reasonable changes in these parameters.

These results are strikingly different from those obtained using the Geneva code.

The use of a custom prescription for rotational mixing during the MS, calibrated to emulate the Geneva code \citep[][Sect.~\ref{ssec:martinelli}]{Martinelli-2025}, although based on calculations for a single track, indicates that more efficient mixing should elevate the luminosity at subsequent evolutionary phases, thus making the results more consistent with those obtained with the Geneva code.

Which of the two predictions for classical Cepheids -- based on the advective–diffusive approach, or on the fully diffusive one -- offers a better match to observations? Clearly, the Geneva tracks have several assets. The strongest, as it can be directly confronted with observations, is the prediction of reasonable, small rotation velocities during core helium burning. The fact that, in this approach, the mass discrepancy may be alleviated is also appealing, but in the authors' opinion does not settle the matter. This is partly due to the degeneracy between the effects of MS core overshooting and rotation-induced mixing. There may also be additional factors contributing to the picture, such as pulsation driven mass loss \citep[][which may also help reduce surface rotational velocities in fully diffusive approach]{Neilson-2012}. 

Surface abundances are another important constraint on the models, in particular on the mixing efficiencies, although they require special attention, as many factors may influence both the abundances and their measurements \citep[see e.g.,][]{Jofre-2019}. Rotation-induced mixing is expected to alter the surface abundances starting already on the MS and later contributing to the first dredge-up on the RGB. A detailed discussion may be found, e.g., in \cite{Choi-2016} for MIST models computed with \MESA{} and in \cite{Ekstrom-2012} for Geneva models. While Geneva models predict more efficient mixing and stronger rotationally induced enhancement of surface abundances, \cite{Anderson-2014} noted that the enhancement of CNO elements is about 0.2\,dex too high. \cite{Takeda-2013} found that CNO abundances in Cepheids are consistent with first dredge-up enhancement as predicted by canonical (non-rotating) evolutionary models. \cite{Smiljanic-2018} analyzed abundances in 32 MW Cepheids and concluded that models with rotation overestimate mixing effects on the surface abundances of C, N, and Na. Their comparison included rotating tracks computed with the advective-diffusive scheme, i.e., tracks computed with the Geneva code \citep{Ekstrom-2012} and with STAREVOL \citep{Lagarde-2012}.

Better observational constraints are clearly needed to resolve the mass discrepancy and examine different implementations of rotation. In addition to accurate measurements of CNO abundances, reliable constraints on period change rates would provide information on the crossing number. Measurements of surface rotational velocities, and their dependence on pulsation period, mass, and metallicity, as well as constraints on mass-loss rates, are needed to probe angular momentum transport. Finally, well-determined physical parameters of Cepheids for a large sample covering a wide range of masses and metallicities are necessary to test the evolutionary models.

The most challenging problem for any implementation of rotation is the ability to reproduce the properties of stars throughout their entire evolution. None of the currently available implementations reproduces all observational constraints. In particular, all codes fail to reproduce the internal rotation of red giants, which can now be well constrained with asteroseismology \citep[e.g.,][]{Aerts-2019-AMRev}. The advective–diffusive scheme is generally considered to provide a more physically consistent description of rotation-induced mixing. While calibrations such as that presented by \cite{Martinelli-2025} may be useful to study specific problems, implementation of the advective–diffusive scheme in \MESA{} appears to be a desirable step forward.


\begin{acknowledgments}
This research is supported by the National Science Center, Poland, Sonata BIS project 2018/30/E/ST9/00598. We thank the referee for a careful reading of the manuscript and for helpful comments. Discussion of \MESA{} rotation settings with Pablo Marchant and Amedeo Romagnolo are acknowledged. We are thankful to Lorenzo Martinelli for discussion about custom rotational-mixing prescription adopted in his paper and to Sylvia Ekstr\"om for clarification about angular momentum transport in the Geneva code. The authors acknowledge the use of ChatGPT (OpenAI) for assistance with improving the grammar and clarity of the manuscript. The authors take full responsibility for the scientific content and conclusions.
\end{acknowledgments}




\software{\MESA{} \citep{Paxton-2011,Paxton-2013,Paxton-2015,Paxton-2018,Paxton-2019,Jermyn-2023}, \texttt{MESA} SDK \citep{Townsend-2022}, modified rotational mixing prescription \citep{Martinelli-2025}, Numpy \citep{Harris-2020}.}

\appendix

\section{Compatibility with newer version of \MESA{}}\label{secapp:oldvsnew}

To demonstrate that our results fully hold with the recent release version of \MESA, 25.12.1, we used it to compute a limited number of models of \set{O14\_ML4h\_RA} set ($\fcor=0.01$, $\fenv=0.04$, $\etar=0.4$) with $W_0=0.5$. We emphasize that an exact one-to-one comparison using an identical physical setup is not possible. This limitation arises from the redefinition of the nuclear reaction networks in the newer version of \MESA, where the option to select reaction rate preferences (REACLIB vs.\ NACRE) is no longer available. Nevertheless, both the previous and updated calculations rely predominantly on reaction rates from REACLIB and employ the same rates for \npg{} \citep{jina} and \cag{} \citep{Kunz-2002}.

\begin{figure*}
\includegraphics[width=\linewidth]{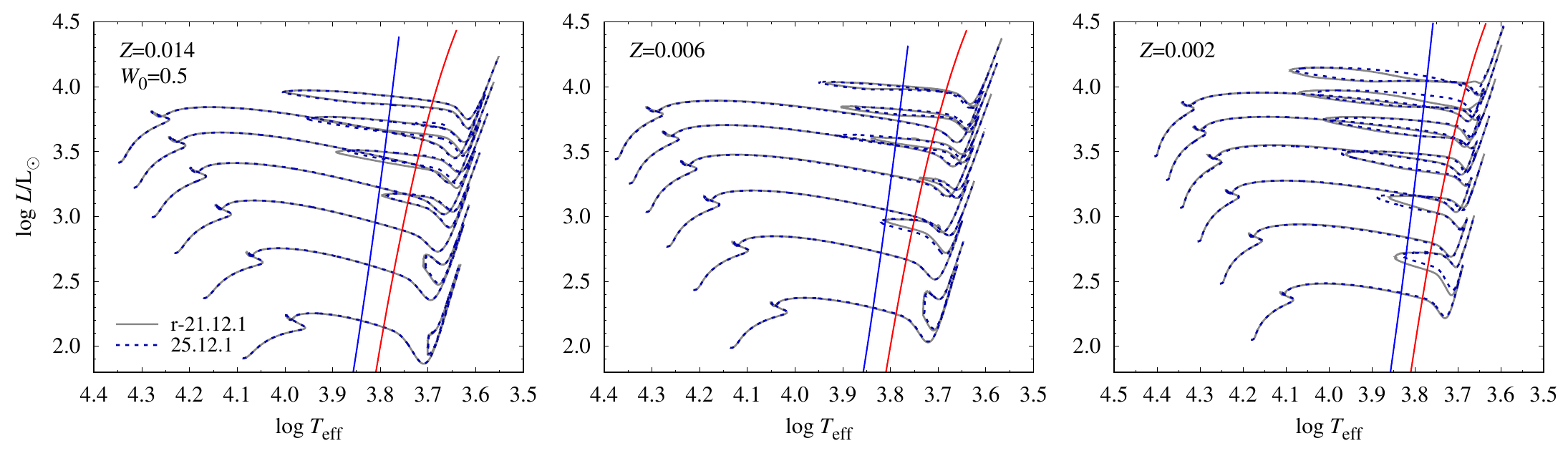}
\caption{Evolutionary tracks computed with \MESA{}-r21.12.1 (solid gray lines) and \MESA-25.12.1 (dotted blue lines) from ZAMS until AGB, for models of 3, 4, 5, 6, 7, and 8\MS{}, and different metallicities, $Z=0.014$ (left), $Z=0.006$ (middle), and $Z=0.002$ (right).
\label{fig:oldnew}}
\end{figure*}

The comparison is illustrated in the HRDs in Fig.~\ref{fig:oldnew}. We observe a very satisfactory agreement of tracks computed with both \MESA{} versions. While some differences are noticeable, the tracks are qualitatively the same. MS tracks follow the same path. Slight differences are present during core helium burning, but the extents of the loops and their luminosity levels are essentially the same. Only for $Z=0.006$, 6\MS{} model, the loop computed with newer \MESA{} version is significantly longer. Still, it crosses through the full IS in both models. We also note that for this specific metallicity the tracks are in general more sensitive e.g., to parameters that control numerical resolution, see Sect.~\ref{secapp:convergence}. Differences in luminosity at core helium burning are typically well below $0.02$\,dex in $\log L/\LS$, and only for 3\MS, $Z=0.002$ model on the 2nd crossing are around $0.04$\,dex. We find that the agreement between the tracks, including the rotation, calculated using the latest \MESA{} version, and the older version used in this study, is very satisfactory.

\section{Effects of core-helium overshooting}\label{secapp:heov}

\begin{figure*}
\includegraphics[width=\linewidth]{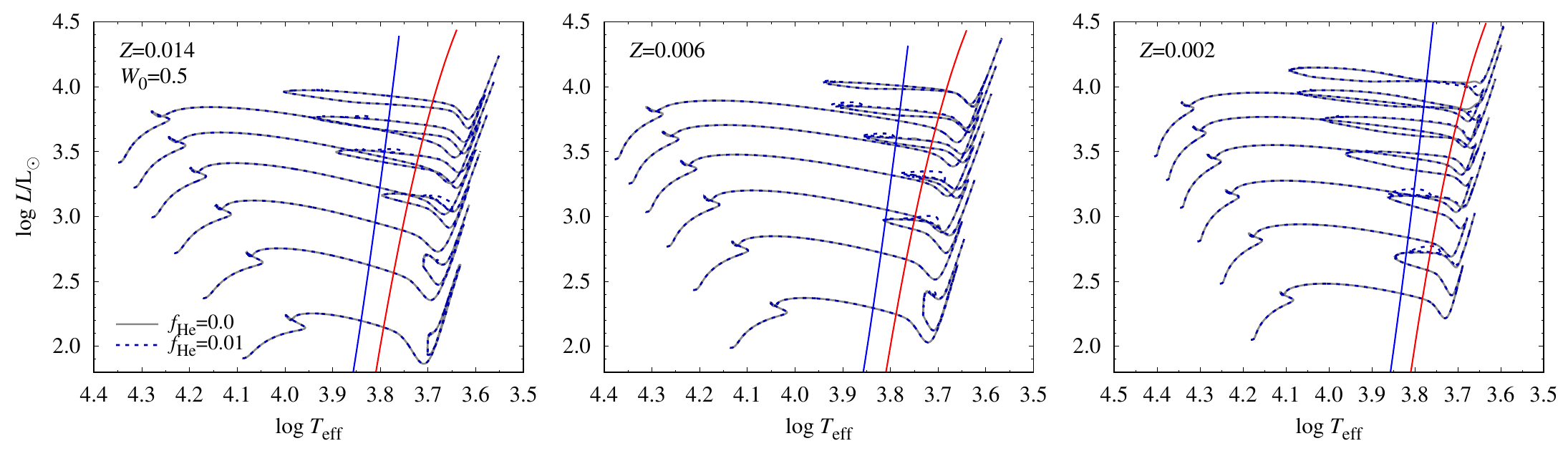}
\caption{Evolutionary tracks computed without core-helium overshooting (\set{O14\_ML4h\_RA} model set, solid gray lines) and including core-helium overshooting (\fhe=0.01, dotted blue lines) from ZAMS until AGB, for models of 3, 4, 5, 6, 7, and 8\MS{}, and different metallicities, $Z=0.014$ (left), $Z=0.006$ (middle), and $Z=0.002$ (right). In all models $W_0=0.5$.
\label{fig:heov}}
\end{figure*}

In Fig.~\ref{fig:heov} we demonstrate the effects of core-helium overshooting on evolutionary tracks including rotation ($W_0=0.5$). Solid gray tracks correspond to models of the \set{O14\_ML4h\_RA} set, i.e., with $\fcor=0.01$, $\fenv=0.04$, $\etar=0.4$, but without convective core overshooting during core-helium burning. In models plotted with dashed blue lines, core-helium overshooting is included with the same efficiency as during the MS evolution, $\fhe=0.01$. Differences between the tracks are very small and are necessarily limited to the blue loop evolution. We observe that the overall extent of the blue loop and the luminosity level during the 2nd and 3rd crossings are essentially the same for models with and without core-helium overshooting. The only qualitative difference is the appearance of secondary loops in a few tracks during the redward evolution toward the AGB. These result from injections of helium into the core due to a less smooth evolution of the core boundary compared to models that do not include core-helium overshooting.

The range of surface rotational velocities during the crossings of the IS is the same in models with and without core-helium overshooting. We conclude that the inclusion of core-helium overshooting in the rotating models does not influence the model properties in any significant way.

\section{Convergence study}\label{secapp:convergence}

In our convergence experiments, we decided to control overall spatial and temporal resolution of the models using the \mdc{} (spatial resolution) as well as \tdc{} (temporal resolution) controls. In addition, for higher temporal resolution, we found it more convenient to decrease the time step by forcing a smaller separation of consecutive models in the HRD. This is achieved with \dhr, \dhrt{} and \dhrl{} controls. We also found it necessary to set \smp, which sets the number of cells used in weighted smoothing of the composition gradient, to zero (no smoothing; the default is 3), as it does not scale linearly with other resolution controls.

\begin{figure*}
\includegraphics[width=\linewidth]{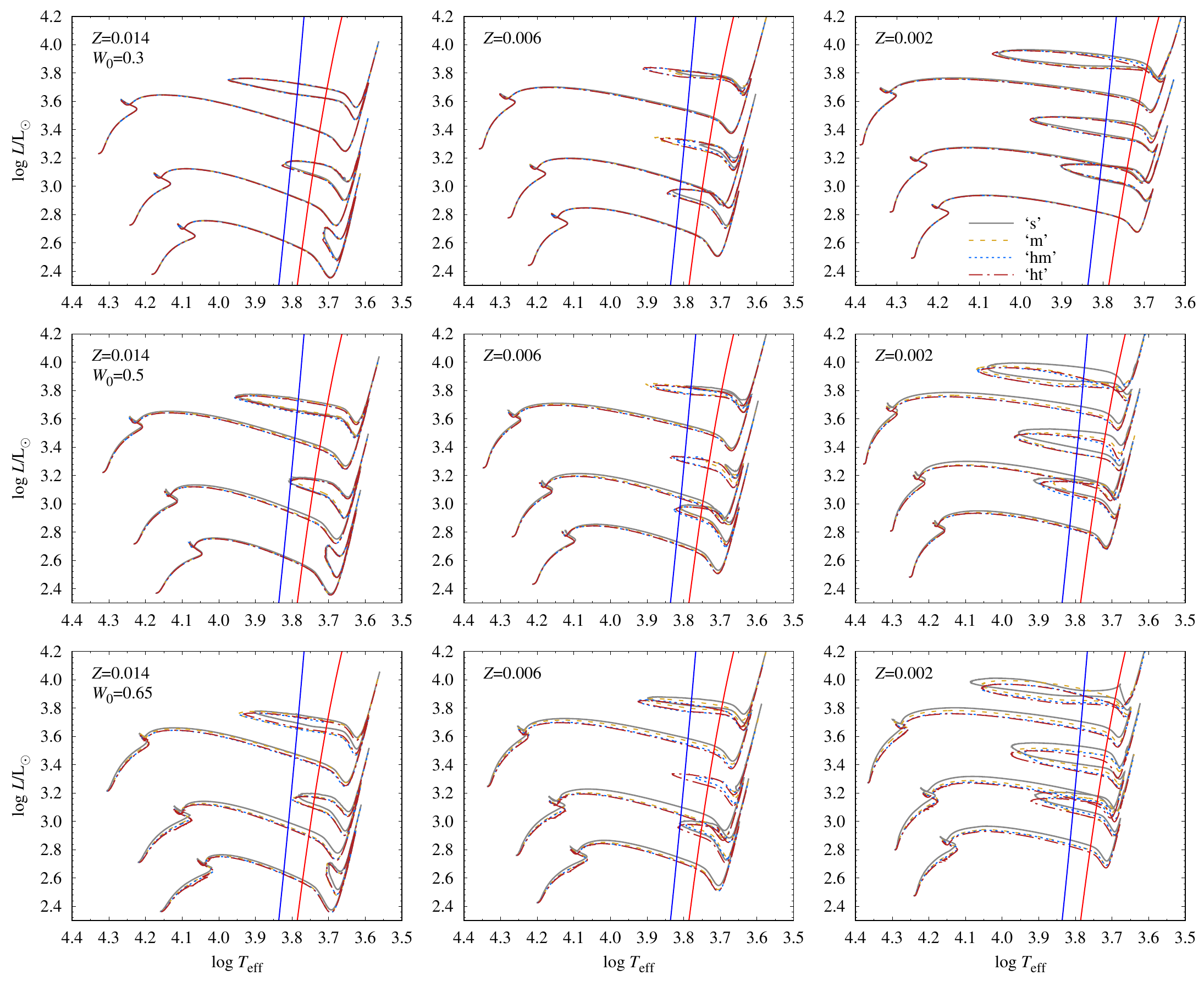}
\caption{Evolutionary tracks computed with different numerical (spatial and temporal) resolution settings for rotating models of 3, 5, and 7\MS{} from ZAMS until AGB. The legend is displayed in the top right panel and numerical controls are given in Tab.~\ref{tabapp:resol}. Models adopt different metallicities, $Z=0.014$ (left), $Z=0.006$ (middle), and $Z=0.002$ (right) and different initial rotation rates, $W_0=0.3$ in the top row, $W_0=0.5$ in the middle row, and $W_0=0.65$ in the bottom row. 
\label{figapp:conv}}
\end{figure*}

Convergence study was conducted assuming overshooting parameters \fcor=0.01 and \fenv=0.04, and $\etar=0.4$, i.e., parameters of model set \set{O14\_ML4h\_RA}, for three metallicities, and three initial rotation rates. As computation of the highest resolution models is very time consuming, we limit the masses to 3, 5, and 7\MS. Four sets of resolution controls were used, as detailed in Tab.~\ref{tabapp:resol}. The first, denoted `s', corresponds to the smallest resolution. Still it is significantly higher then the \MESA{} default. This is also the set of resolution controls we have used for non-rotating models in \citetalias{Ziolkowska-2024}--\citetalias{Smolec-2026} and as we show in \citetalias{Ziolkowska-2024} the results are then converged. In the medium resolution set, `m', the \mdc{} and \tdc{} controls are cut by two. In `hm' set we use even higher spatial resolution, and in `ht', in addition to higher spatial resolution we also increase the temporal resolution (the total number of models in evolutionary tracks until the end of core helium burning is typically about 60\% larger in `ht' model as compared with `hm' model). The resulting tracks are presented in Fig.~\ref{figapp:conv}.

\begin{deluxetable}{lrrrr}
\tablewidth{0pt}
\tablecaption{Spatial and temporal resolution controls and their values adopted in four model sets, `s', `m', `hm' and `ht', considered in convergence study. \label{tabapp:resol}}
\tablehead{
\colhead{\MESA's inlist control} & \colhead{`s'} & \colhead{`m'} & \colhead{`hm'} & \colhead{`ht'}}
\startdata
\mdc  & 0.5   & 0.25  & 0.125 & 0.125\\
\tdc  & 0.5   & 0.25  & 0.25  & 0.25\\
\dhr  & 0.005 & 0.005 & 0.005 & 0.0025\\
\dhrt & 0.005 & 0.005 & 0.005 & 0.0025\\
\dhrl & 0.01  & 0.01  & 0.01  & 0.005\\
\enddata
\end{deluxetable}

We observe that when using the lowest resolution set, `s', which guaranteed convergence for non-rotating models, the convergence is no longer present for rotating models, which may be noted already for models with the smallest initial rotation rate. We find that overall luminosity of the models is in general larger for the smallest resolution models. For three other sets with increased spatial and temporal resolution, consistency between the tracks is significantly higher. Differences increase with increasing initial rotation rate. For MS models, we observe qualitatively and quantitatively the same tracks. Differences increase with evolution stage. For core helium burning and $Z=0.014$ and $Z=0.002$ models, we note that the three sets of higher resolution, `m', `hm', and `ht', result in qualitatively the same tracks, i.e., the loop extent and luminosity levels are very similar. Only for intermediate metallicity models, $Z=0.006$, qualitative differences are recorded. These concern e.g., the extent of the loops, which may be significantly different. For $W_0=0.65$ and 5\MS{} model, the loop does not develop in `m' models, then it reaches the red edge of the IS for `hm' models, and extends slightly beyond blue edge in `ht' models.

Apart from the lowest resolution models, the differences between evolutionary tracks, although increasing with increasing initial rotation rate, are small. While for some models the differences may be qualitative, considering the computation time, the set of parameters with medium resolution, `m', seems to be a reasonable compromise ensuring good convergence over a wide range of parameters, masses, metallicities, and initial rotation rates, and this set has been adopted as the default for all computations in this work.

\section{Evolutionary period change rates}\label{secapp:pcr}

\begin{figure*}
\includegraphics[width=\linewidth]{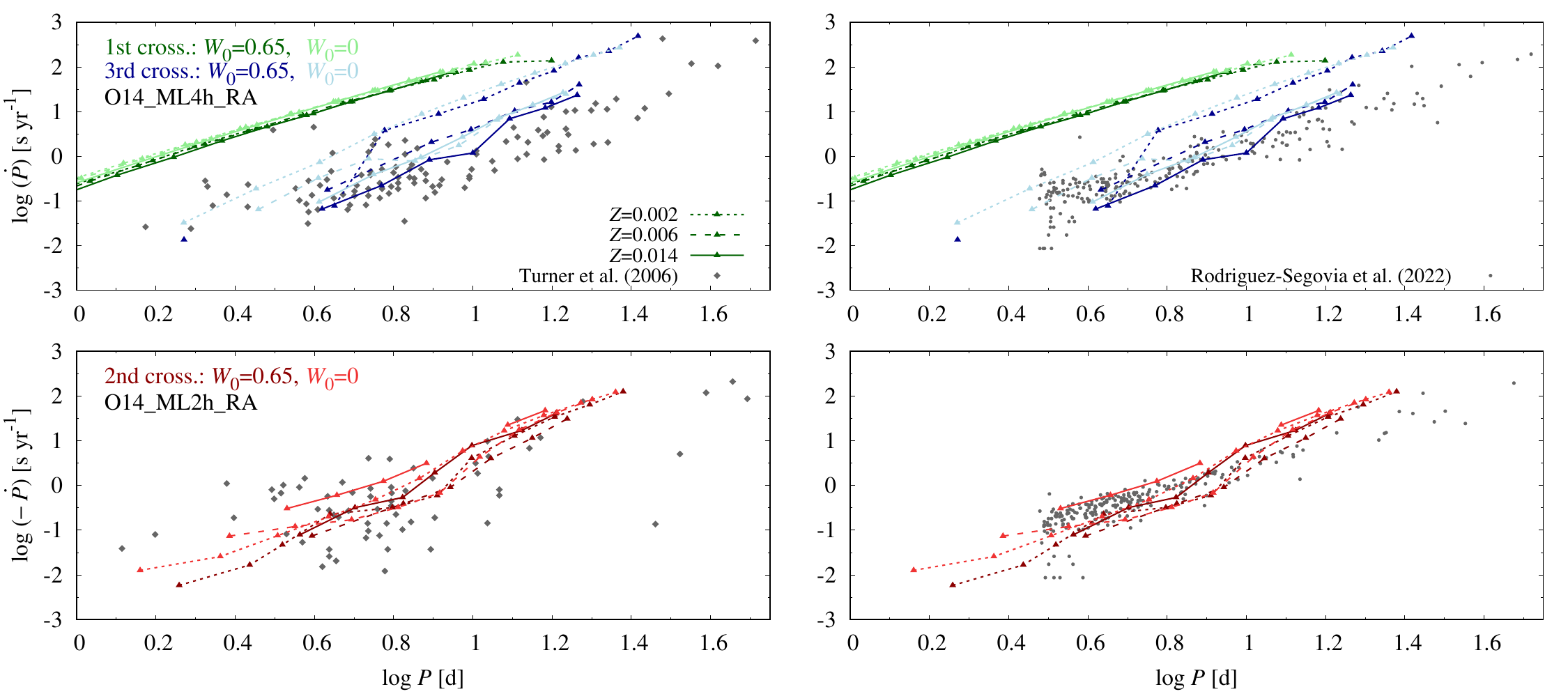}
\caption{Positive (top) and negative (bottom) period change rates predicted form evolutionary models of different metallicities (different line styles), including rotation (\set{O14\_ML4h\_RA}, $W_0=0.65$, dark lines) and non-rotating (bright lines), confronted with determinations for MW Cepheids from \cite{Turner-2006} and LMC Cepheids from \cite{Rodriguez-Segovia-2022}. PCRs were avaluated at the midline of the IS, separately for each crossing.
\label{figapp:pcr}}
\end{figure*}

In Fig.~\ref{figapp:pcr} we show evolutionary period change rates (PCRs) for models of \set{O14\_ML4h\_RA} set, assuming the largest initial rotation rate, $W_0 = 0.65$, compared with PCRs derived from non-rotating tracks and with observational data for MW Cepheids \citep[][left panels]{Turner-2006} and LMC Cepheids \citep[][right panels]{Rodriguez-Segovia-2022}. It is evident that the differences between predictions based on rotating and non-rotating models are rather small, even for the largest rotation rate considered in this study. PCRs based on rotating models are, on average, slightly lower (i.e., the evolution is slower). In general, the predictions are consistent with observations. Some discrepancies noted in \citetalias{Smolec-2026} (see Sect.~3.6) are not resolved by the inclusion of rotation. We note, however, that a meaningful comparison with observations requires population synthesis, which is beyond the scope of this paper. The presented results and trends are consistent with those discussed by \cite{EspinozaArancibia-2022}.

In Tab.~\ref{tabapp:agextime} we provide an analog to Tab.~6 in \citetalias{Smolec-2026}, giving crossing times through the IS and period change rates expressed as $\dot{P}/P$ (yr$^{-1}$), evaluated at the midline of the IS, separately for each crossing and for all model sets included in the top section of Tab.~\ref{tab:models}. Only a sample of the table is included for reference. The full table is available in the electronic edition of the Journal and on Zenodo.

\begin{deluxetable*}{lccclllllllll}
\tablecaption{Logarithm of age (yrs) at the entry of the IS, logarithm of crossing time (yrs), $\log t_{\rm cross}$, and period change rate, $\dot{P}/P$ (yr$^{-1}$; at the midpoint of the IS) for the three IS crossings. The first four columns identify model set, initial angular rotation rate, $W_0$, metallicity and initial mass, respectively. For the 2nd and 3rd crossing, asterisk at crossing time indicates the tip of the loop falls inside the IS (and the crossing time is computed based on the model's age at the tip of the loop and respective edge of the IS). \label{tabapp:agextime}}
\tablehead{
  &   &     &     & \multicolumn{3}{c}{ 1st crossing} & \multicolumn{3}{c}{2nd crossing} & \multicolumn{3}{c}{3rd crossing}\\
\colhead{set} & \colhead{$W_0$} & \colhead{$Z$} & \colhead{$M_0/\MS$} & \colhead{log age} & \colhead{$\log t_{\rm cross}$} & \colhead{$\dot{P}/P$} & \colhead{log age} & \colhead{$\log t_{\rm cross}$} & \colhead{$\dot{P}/P$}  & \colhead{log age} & \colhead{$\log t_{\rm cross}$}  & \colhead{$\dot{P}/P$}
}
\startdata
O00\_ML4h\_RA & 0.00 & 0.0140 & 2.5 &    8.7040  & 6.0482  & 2.4263e-07 &          -  &      -  &           - &          -  &      -  &          - \\
O00\_ML4h\_RA & 0.00 & 0.0140 & 3.0 &    8.4924  & 5.5167  & 7.6793e-07 &          -  &      -  &           - &          -  &      -  &          - \\
O00\_ML4h\_RA & 0.00 & 0.0140 & 3.5 &    8.3183  & 5.1485  & 1.7842e-06 &          -  &      -  &           - &          -  &      -  &          - \\
O00\_ML4h\_RA & 0.00 & 0.0140 & 4.0 &    8.1708  & 4.8701  & 3.5005e-06 &          -  &      -  &           - &          -  &      -  &          - \\
O00\_ML4h\_RA & 0.00 & 0.0140 & 4.5 &    8.0440  & 4.6468  & 6.1868e-06 &     8.1245  & 6.1571* & -1.3089e-07 &     8.1292  & 6.3651* & 1.3283e-07 \\
\multicolumn{13}{l}{\ldots}\\
O00\_ML4h\_RA & 0.30 & 0.0140 & 4.5 &    8.0567  & 4.6479  & 6.2284e-06 &     8.1235  & 6.2646* &           - &     8.1295  & 6.5245* &          - \\
O00\_ML4h\_RA & 0.30 & 0.0140 & 5.0 &    7.9467  & 4.4586  & 1.0315e-05 &     8.0167  & 6.3987* & -1.4567e-07 &     8.0271  & 6.2393* & 2.8756e-07 \\
O00\_ML4h\_RA & 0.30 & 0.0140 & 5.5 &    7.8522  & 4.3026  & 1.5935e-05 &     7.9172  & 6.1007  & -3.2874e-07 &     7.9342  & 5.8218  & 8.4870e-07 \\
\multicolumn{13}{l}{\ldots}\\
\enddata
\tablecomments{This table is published in its entirety in the electronic edition of the {\it Astrophysical Journal Supplement Series} and on Zenodo. A portion is shown here for guidance regarding its form and content.}
\end{deluxetable*}

\clearpage

\bibliography{paper2_arXiv}{}
\bibliographystyle{aasjournal}

\end{document}